\documentclass[lettersize,journal]{IEEEtran}
\usepackage{amsmath,amsfonts}
\usepackage{algorithmic}
\usepackage{algorithm}
\usepackage{array}
\usepackage[caption=false,font=normalsize,labelfont=sf,textfont=sf]{subfig}
\usepackage{textcomp}
\usepackage{stfloats}
\usepackage{url}
\usepackage{verbatim}
\usepackage{graphicx}
\usepackage{cite}
\usepackage{bm}
\usepackage{colortbl}
\usepackage{mathtools}
\usepackage{arydshln}
\usepackage{booktabs}
\usepackage{multicol}
\usepackage{multirow}
\usepackage{hhline}
\usepackage{makecell}
\usepackage{color}
\usepackage{xcolor}
\usepackage{orcidlink}

\usepackage{enumitem}
\definecolor{defblue}{RGB}{0,120,191}
\definecolor{defred}{RGB}{204,64,84}
\definecolor{defgray}{RGB}{229,245,252}
\definecolor{defgreen}{RGB}{0,128,0}
\begin{document}

\title{Self-Training Boosted Multi-Factor Matching Network for Composed Image Retrieval}

\author{Haokun~Wen~\orcidlink{0000-0003-0633-3722},
        Xuemeng~Song~\orcidlink{0000-0002-5274-4197}\IEEEmembership{, Senior Member, IEEE},
        Jianhua~Yin~\orcidlink{0000-0002-4611-2986}\IEEEmembership{, Member, IEEE},         
        Jianlong~Wu~\orcidlink{0000-0003-0247-5221 }\IEEEmembership{, Member, IEEE}, 
        Weili~Guan~\orcidlink{0000-0002-5658-5509}\IEEEmembership{, Member, IEEE},
        Liqiang~Nie~\orcidlink{0000-0003-1476-0273}\IEEEmembership{, Senior Member, IEEE}

\thanks{This work was supported in part by the National Natural
Science Foundation of China under Grants 62236003, 62376137, 62172261, and 62376069, in part by the Shandong Provincial Natural Science Foundation under Grants ZR2022YQ59, and in part by the Young Elite Scientists Sponsorship Program by CAST under Grants 2023QNRC001. Haokun Wen acknowledges the support from the China Scholarship Council for the completion of this work at the National University of Singapore.}
\thanks{Haokun Wen, Jianlong Wu, and Liqiang Nie are with the School of Computer Science and Technology, Harbin Institute of Technology (Shenzhen), 518055, China. E-mail: whenhaokun@gmail.com, jlwu1992@pku.edu.cn, nieliqiang@gmail.com.}
\thanks{Xuemeng Song and Jianhua Yin are with the School of Computer Science and Technology, Shandong University, Tsingtao, 266000, China. E-mail: sxmustc@gmail.com, jhyin@sdu.edu.cn.}
\thanks{Weili Guan is with the School of Electronics and Information Engineering, Harbin Institute of Technology (Shenzhen), 518055, China. E-mail: honeyguan@gmail.com.}
\thanks{Xuemeng Song and Liqiang Nie are the corresponding authors.}

}

\markboth{IEEE TRANSACTIONS ON PATTERN ANALYSIS AND MACHINE INTELLIGENCE,~Vol.~16, No.~5, May~2024}
{Wen \MakeLowercase{\textit{et al.}}: Self-Training Boosted Multi-Factor Matching Network for Composed Image Retrieval}


\maketitle

\begin{abstract}
The composed image retrieval (CIR) task aims to retrieve the desired target image for a given multimodal query, \textit{i.e.}, a reference image with its corresponding modification text. The key limitations encountered by existing efforts are two aspects: 1) ignoring the multiple \mbox{query-target} matching factors; 2) ignoring the potential unlabeled \mbox{reference-target} image pairs in existing benchmark datasets. To address these two limitations is non-trivial due to the following challenges: 1) how to effectively model the multiple matching factors in a latent way without direct supervision signals; 2) how to fully utilize the potential unlabeled \mbox{reference-target} image pairs to improve the generalization ability of the CIR model. To address these challenges, in this work, we first propose a CLIP-Transformer based \mbox{muLtI-factor} Matching Network (LIMN), which consists of three key modules: disentanglement-based latent factor tokens mining, dual aggregation-based matching token learning, and dual query-target matching modeling. Thereafter, we design an iterative dual \mbox{self-training} paradigm to further enhance the performance of LIMN by fully utilizing the potential unlabeled \mbox{reference-target} image pairs in a \mbox{weakly-supervised} manner. Specifically, we denote the iterative dual \mbox{self-training} paradigm enhanced LIMN as LIMN+. Extensive experiments on four datasets, including FashionIQ, Shoes, CIRR, and Fashion200K, show that our proposed LIMN and LIMN+ significantly surpass the state-of-the-art baselines.
\end{abstract}


%
%
%
%
\section{Introduction}
\IEEEPARstart{I}{mage} retrieval has been a cardinal task in computer vision and database management domains since the $1970$'s~\cite{ir}. It has laid the foundation and paved the way for a wide range of applications, such as face recognition~\cite{face1,face2}, fashion retrieval~\cite{fashion1}, and person re-identification~\cite{reid1, reid2}. Traditional image retrieval paradigms allow users to deliver their search intentions by either pure text queries or image queries. However, in many cases, the user's search intention is sophisticated, and cannot be well delivered by the \mbox{single-modal} query. Accordingly, to facilitate the flexible expression of users' search intentions, Vo \textit{et al.}~\cite{tirg} first proposed the task of composed image retrieval (CIR), where the input query is composed of a reference image and a modification text describing the user's modification demands towards the reference image. As CIR has great potential value in many \mbox{real-world} applications, such as commercial product search and interactive intelligent robots, it has gained increasing research attention in recent years.

\begin{figure}[t]
	\includegraphics[width=0.98\linewidth]{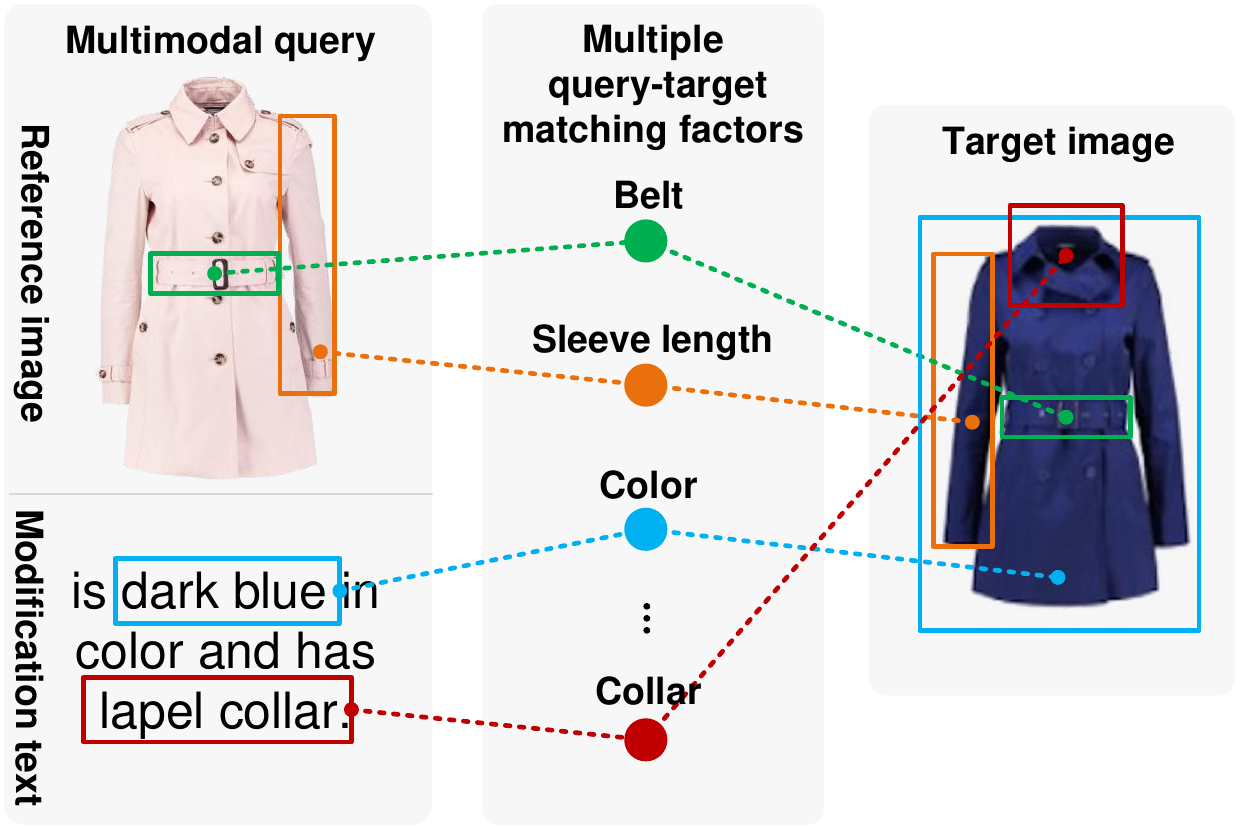}
	\caption{An example of the multiple query-target matching factors between the \mbox{multimodal} query and the target image. Note that semantics like ``color'' or ``belt'' only convey a concept since the matching factors are implicitly contained.}
    \vspace{-0.5em}
	\label{fig:example}
\end{figure}

Although the pioneer research studies~\cite{val,yangsigir,artemis,clvcnet,tgcir,composed,cosmo,nietip} have achieved promising progress, they have two key limitations. 
\begin{itemize}[leftmargin=17pt]
    \item[\textbf{L1:}] \textbf{Ignore the multiple query-target matching factors.} As shown in Figure~\ref{fig:example}, for evaluating whether the target image matches the multimodal query, we usually need to consider multiple factors (\textit{e.g.}, the color, collar, and sleeve length). Nevertheless, none of the existing efforts take this into consideration. The previous studies simply represent the multimodal query and the target image by a single overall vector, respectively, based on which they conduct the \mbox{query-target} matching. Apparently, this strategy fails to clearly model the various latent factors affecting the query-target matching evaluation and inevitably leads to suboptimal performance.
    \item[\textbf{L2:}] \textbf{Ignore the potential unlabeled reference-target image pairs in the existing benchmark dataset.}
    Samples in the existing CIR datasets are all in the form of triplet \textless reference image, modification text, target image\textgreater. In particular, the reference-target image pairs are obtained in a heuristic manner, while the modification text is human-annotated.
    Apparently, creating such training samples is expensive and laborious, which largely limits the size of the benchmark datasets. Simply relying on the limited samples in existing benchmark datasets, previous research efforts encountered the overfitting problem to some extent and suffered from poor generalization ability. 
    In fact, we observed that there are still plentiful potential unlabeled \mbox{reference-target} image pairs in the dataset being untapped by previous work, which are highly similar with few different properties.
\end{itemize}

Overcoming the flaws mentioned above is \mbox{non-trivial} due to the following two challenges. 
1) Each factor that affects the query-target matching evaluation should emphasize its correlated features of the multimodal query and the target image. Meanwhile, we do not have direct supervision signals for learning the multiple \mbox{query-target} matching factors. Therefore, how to effectively model the multiple \mbox{query-target} matching factors in a latent way constitutes the first challenge.
2) Although a number of potential unlabeled \mbox{reference-target} image pairs can be heuristically uncovered from the benchmark dataset, they cannot be used directly for training the CIR model since they lack the corresponding modification text.  Accordingly, another tough challenge is how to fully utilize the potential unlabeled \mbox{reference-target} image pairs in a \mbox{weakly-supervised} manner to improve the generalization ability of the CIR model.

To address the challenges mentioned above, we first propose a CLIP-Transformer based muLtI-factor Matching Network, LIMN for short, to model the multiple \mbox{query-target} matching factors between the multimodal query and the target image. Thereafter, we devise a customized iterative dual \mbox{self-training} paradigm to make full use of the potential unlabeled \mbox{reference-target} image pairs and solve the task in a \mbox{weakly-supervised} manner. We name the iterative dual self-training paradigm enhanced LIMN as LIMN+.
 
For LIMN, it contains three vital modules: disentanglement-based latent factor tokens mining, dual aggregation-based matching token learning, and dual query-target matching modeling. The first module is dedicated to disentangling the CLIP-based global features of the multimodal query and the target image, paving the way for extracting multiple latent factor tokens. In the second module, we utilize the masked Transformer~\cite{transformer} to attentively aggregate the local features into the multiple latent factor tokens. Meanwhile, we aggregate these tokens into the final matching token, for ensuring both effectiveness and efficiency during the inference stage. In the third module, we employ the commonly used \mbox{batch-based} classification loss for making the matching tokes of the multimodal query and the target image close. Besides, we also apply an explicit constraint on the intermediate latent factor tokens by extending the batch-based classification loss to prompt the multi-factor matching modeling.

As for LIMN+, the key novelty lies in our proposed iterative dual \mbox{self-training} paradigm. It takes full advantage of the existing image difference captioning (IDC)~\cite{acl2018} model that can generate a natural \mbox{language-based} text to describe the difference between two images. Specifically, our iterative dual \mbox{self-training} paradigm utilizes the IDC model to automatically annotate the potential unlabeled \mbox{reference-target} image pairs and produce pseudo triplets for enhancing the training of the LIMN model. Conversely, the \mbox{well-trained} LIMN model can measure the quality of the pseudo triplet by evaluating its \mbox{query-target} matching score and determine whether the pseudo triplets can be used. Through iterative cooperation learning, the performance of LIMN gets improved.

Our main contributions can be summarized as follows:
\begin{itemize}
    \item As far as we know, we are the first to explore the latent multiple query-target matching factors in the context of CIR. In particular, we propose a CLIP-Transformer based \mbox{multi-factor} matching network, which first mines the latent query-target matching factor tokens with global feature disentanglement and then learns the final matching token based on attentive local feature aggregation and latent factor aggregation.
    \item  To the best of our knowledge, we are the first to dissect the scarce dataset problem encountered by the CIR task. Specifically, we \mbox{design} an iterative dual \mbox{self-training} paradigm to make full use of the potential unlabeled \mbox{reference-target} image pairs and further boost the CIR performance. The paradigm is \mbox{hot-plugging} and is applicable to any other existing CIR methods. 
    \item Extensive experiments on four datasets demonstrate the superiority of our proposed models (LIMN and LIMN+). As a byproduct, we released the source code to facilitate the research community\footnote{\url{https://anosite.wixsite.com/limn}.}.
\end{itemize}

The remainder of the paper is organized as follows. Section $2$ briefly reviews the related work. In Section $3$, we detail the proposed LIMN and iterative dual self-training paradigm. The experimental results and analyses are presented in Section $4$, followed by the conclusion and future work in Section $5$.

\section{Related Work}
Our work is closely related to the studies on composed image retrieval (CIR) and self-training.

\subsection{Composed Image Retrieval}

Recent years have witnessed growing research interest in CIR due to its theoretical significance of research and potential commercial value.
There are various efforts~\cite{tirg,val,composeae,clvcnet,dcnet,artemis,cosmo,SAC,JGAN,crr,lin2023clip,cvpr21,niemm,guo1,guo2} that attempt to tackle this task. For instance, Vo \textit{et al.}~\cite{tirg} first proposed a gating and residual module to compose the multimodal query features, where the reference image is adaptively preserved and transformed to derive the composed query representation. Following that, Chen \textit{et al.}~\cite{val} resorted to the attention mechanism to fuse the hierarchical reference image features with the modification text feature. Later, Kim \textit{et al.}~\cite{dcnet} presented a refined version of TIRG~\cite{tirg} to fuse the \mbox{multi-granularity} features of the reference image and modification text. Thereinto, a correction network on the difference between the reference image and target image is introduced to further regularize the fusion results. Moreover, Wen \textit{et al.}~\cite{clvcnet} utilized a \mbox{fine-grained} \mbox{local-wise} composition module and a \mbox{fine-grained} \mbox{global-wise} composition module, which can model the diverse modification demands more precisely. Thereafter, Lee \textit{et al.}~\cite{cosmo} first adopted a content modulator to perform local updates of the reference image according to the modification text, and then applied a style modulator to achieve the global modification.

Although these methods have made prominent progress, they overlook the multiple matching factors that affect the matching evaluation between the multimodal query and the target image in the context of \mbox{CIR}, and inevitably obtain the suboptimal retrieval performance. In addition, the existing efforts ignore the potential unlabeled \mbox{reference-target} image pairs in the existing benchmark dataset.
Beyond these studies, in this work, we captured the multiple query-target matching factors by extracting multiple matching tokens to achieve promising retrieval performance. Besides, we \mbox{designed} an iterative dual \mbox{self-training} algorithm to make full use of the potential unlabeled \mbox{reference-target} image pairs and relieve the problems caused by limited data size, and consequently achieved promising results in \mbox{CIR}.

\subsection{Self-training}

\mbox{Self-training} is one of the earliest techniques in \mbox{weakly-supervised} learning~\cite{stFralick67}. Typically, by \mbox{self-training}, a model is first trained under the supervision of the labeled data. And then the \mbox{well-trained} model is employed to label the unlabeled data to derive the pseudo labeled data. Finally, both the labeled and the pseudo labeled data are jointly utilized to train another model, which can yield better performance. The advantage of self-training is that it can adapt the supervised model to function in a \mbox{weakly-supervised} manner and learn from the unlabeled data. This strategy is simple yet effective and has drawn many researchers' attention. 
For example, Xie \textit{et al.}~\cite{cvprself1} utilized the \mbox{self-training} technique to significantly advance the image classification accuracy by adding noise such as dropout, stochastic depth and data augmentation to the combination of the labeled and the pseudo labeled data. In addition, Ye \textit{et al.}~\cite{aclself} adopted \mbox{self-training} in \mbox{zero-shot} text classification, where a reinforced learning framework is proposed to select more reliable data samples automatically.
Moreover, Yang \textit{et al.}~\cite{cvprself2} employed \mbox{self-training} in \mbox{weakly-supervised} object detection, in which the non-maximum suppression is utilized to fuse the detection results from different iterations and the performance is further improved. 

Inspired by these successful applications of \mbox{self-training}, in this work, we adapted the \mbox{self-training} technique to the CIR task to alleviate the limited dataset problem. Considering the setting of CIR, we adopted an image difference captioning model to produce pseudo triplets for promoting the performance of the CIR model.

\section{Methodology}
In this section, we first formulate the research problem and then elaborate on the proposed CLIP-Transformer based muLtI-factor Matching Network (LIMN), followed by the iterative dual self-training paradigm.

\subsection{Problem Formulation}
In this work, we aim to tackle the task of CIR, whose goal is to retrieve the target images that meet the multimodal query (\textit{i.e.}, a reference image and a modification text). Suppose we have a set of $N$ training triplets, denoted as $\mathcal{T}=\left\{ \left( x_{r}, t_{m}, x_{t} \right)_{i} \right\}_{i=1}^{N}$, where $x_{r}$, $t_{m}$, and $x_{t}$ refer to the reference image, modification text, and target image, respectively. 
Essentially, we aim to learn an embedding space where the representations of the multimodal query $\left(x_r, t_m\right)$ and corresponding target image $x_t$ should be as close as possible, which can be formally expressed as follows,
\begin{equation}
    \mathcal{H}\left(x_r, t_m\right) \rightarrow \mathcal{H}\left(x_t\right),\label{eq1}
\end{equation}
where $\mathcal{H}$ denotes the \mbox{to-be-learned} mapping function for embedding both the multimodal query and the target image. Here we utilize the same mapping function to ensure the representations of the multimodal query and the target image lying in the same semantic embedding space.

\begin{figure*}[ht]
    \centering
    \includegraphics[width=0.95\textwidth]{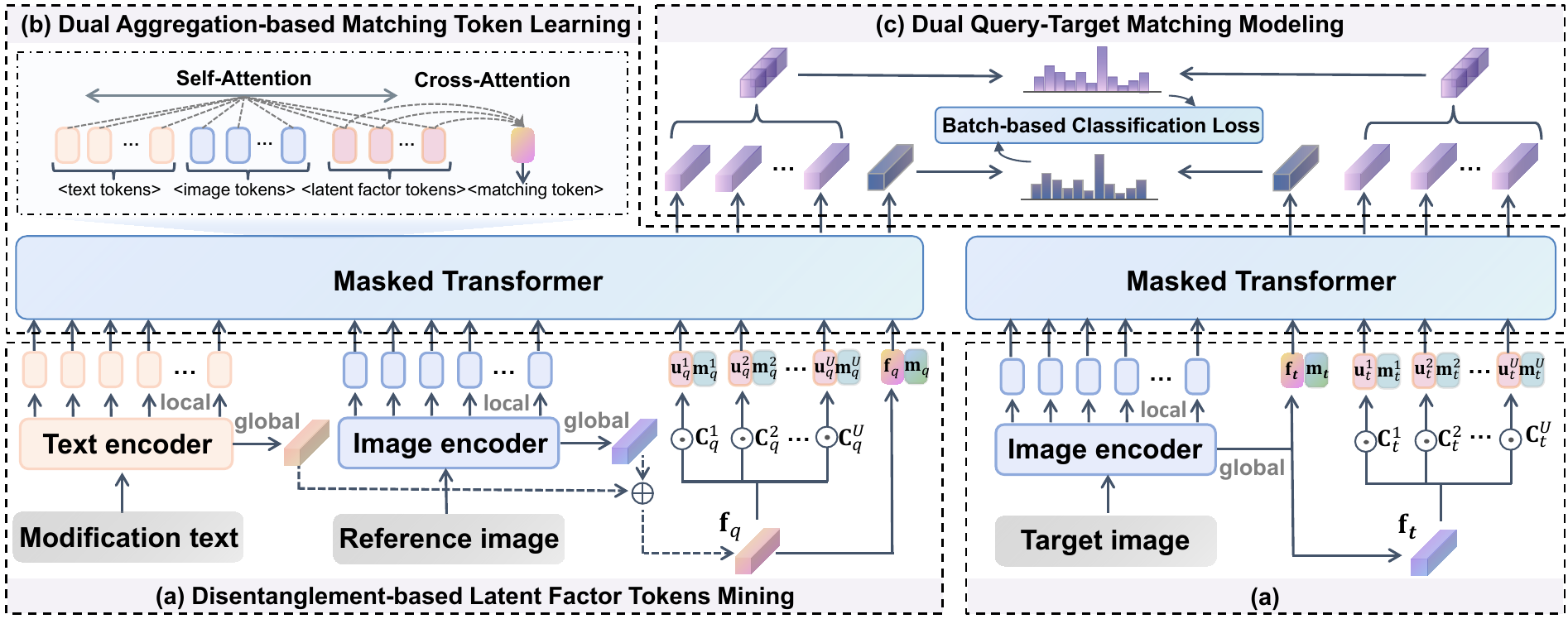}
    \caption{Illustration of our proposed LIMN. It consists of three modules: (a) disentanglement-based latent factor tokens mining, (b) dual aggregation-based matching token learning, and (c) dual query-target matching modeling.}\label{fig:limn}
    \vspace{-1em}
\end{figure*}

\subsection{LIMN}
One major novelty of our work is that we take the multiple factors influencing the matching evaluation between the multimodal query and the target image into account. 
Specifically, we propose a CLIP-Transformer based muLtI-factor Matching Network (LIMN). As shown in Figure~\ref{fig:limn}, the proposed LIMN consists of three modules.
(a) Disentanglement-based latent factor tokens mining, which can uncover the latent factor tokens via disentangling the CLIP-based global query/target features (detailed in Section 3.2.1).
(b) Dual aggregation-based matching token learning, which jointly conducts the attentive local feature aggregation and attentive latent factor aggregation with a masked Transformer, to learn the final matching token to-be-used in the inference stage (explained in Section 3.2.2).
And (c) dual query-target matching modeling, which deploys the conventional ranking loss over both the final matching token and latent factor tokens (described in Section 3.2.3). We now elaborate each module in LIMN as follows.

\subsubsection{Disentanglement-based Latent Factor Tokens Mining}
As aforementioned, we do not have explicit supervision signals for multiple latent factors learning. To circumvent this impediment, we propose to learn the latent matching factors based on feature disentanglement. In particular, in this work, we adopt the contrastive language-image pre-training model (CLIP)~\cite{clip} as the feature extraction backbone, which has shown remarkable success in the context of CIR~\cite{clip4cir,tgcir}.
Let $\mathbf{f}_r \in \mathbb{R}^{D}$, $\mathbf{f}_m \in \mathbb{R}^{D}$, and $\mathbf{f}_t \in \mathbb{R}^{D}$ denote the global features of the reference image, modification text, and target image, respectively, all of which are outputted by CLIP's last layer. To facilitate the feature disentanglement for latent factor learning, we first fuse the global features of the multimodal query with the straightforward weighted summation strategy as follows,
\begin{equation}
\left\{\begin{aligned}
\mathbf{w} &= \sigma \left( \operatorname{MLP}\left( \left[ \mathbf{f}_r;\mathbf{f}_m \right] \right) \right),\\
\mathbf{f}_q &= \mathbf{w} \odot \mathbf{f}_r  + \left(1-\mathbf{w}\right) \odot \mathbf{f}_m, 
\end{aligned}\right.
\label{eq-fuse}
\end{equation}
where  $\operatorname{MLP}$ refers to the multi-layer perceptron, $\sigma$ refers to the Sigmoid activation function, $\odot$ is the element-wise multiplication, and $\mathbf{f}_q \in \mathbb{R}^{D}$ denotes the fused representation of the multimodal query.

Suppose there are $U$ query-target matching factors, we then accordingly introduce $U$ learnable condition masks~\cite{zhengnamm} for extracting the latent $U$ factor tokens from both the query and target sides. Formally, we have,
\begin{equation}
\left\{\begin{aligned}
\mathbf{u}_{q}^{i} &= \mathbf{f}_{q} \odot \mathbf{C}_{q}^{i}, \\
\mathbf{u}_{t}^{i} &= \mathbf{f}_{t} \odot \mathbf{C}_{t}^{i},
\end{aligned}\right.
\label{eq-fuse}
\end{equation}
where $\mathbf{C}_{q}^{i} \in \mathbb{R}^{D}$ and $\mathbf{C}_{t}^{i} \in \mathbb{R}^{D}$ are the $i$-th condition masks for the multimodal query and the target image, respectively, with $\mathbf{u}_{q}^{i}$ and $\mathbf{u}_{t}^{i}$ denoting the corresponding latent factor tokens.

To ensure that each latent factor token focuses on a single specific matching factor, we take the following two strategies. On one hand, following~\cite{zhengnamm,mask}, we employ the L$1$ regularization on the condition masks as follows,
\begin{equation}
    \mathcal{L}_{L1} = \frac{1}{U} \sum_{i=1}^{U} \left\| \mathbf{C}_{q}^{i} \right\| + \frac{1}{U} \sum_{i=1}^{U} \left\| \mathbf{C}_{t}^{i} \right\|. \label{eq-mask}
\end{equation}
On the other hand, each latent factor token is associated with a distinct factor identifier token, which is randomly initialized and can be optimized during the training process. Introducing factor identifier tokens can avoid homogenization and promote the independence of latent factor tokens. Mathematically, we denote the final query latent factor tokens as $\mathbf{X}_q = \left[ \mathbf{u}_{q}^{1} + \mathbf{m}_{q}^{1}, \dots,  \mathbf{u}_{q}^{U} + \mathbf{m}_{q}^{U}  \right] \in \mathbb{R} ^{U \times D}$, and final target latent factor tokens as $\mathbf{X}_t = \left[ \mathbf{u}_{t}^{1} + \mathbf{m}_{t}^{1}, \dots,  \mathbf{u}_{t}^{U} + \mathbf{m}_{t}^{U} \right] \in \mathbb{R} ^ {U \times D}$. $\mathbf{m}_{q}^{i}$ and $\mathbf{m}_{t}^{i}$ denote the identifier token for the $i$-th query and target latent factor tokens, respectively.

\subsubsection{Dual Aggregation-based Matching Token Learning}
Intuitively, the latent query-target matching factors may be correlated to certain concrete local features. Suppose there is a latent factor referring to the ``belt'' attribute of a garment. This factor should be particularly interested in some related regions of the query/target image or some related keywords in the modification text. Therefore, we argue that the latent factor tokens learned by global feature-based disentanglement could be enhanced by further aggregating the correlated local features of the query/target. 
Additionally, directly leveraging multiple latent factor tokens to achieve multi-factor matching during inference could potentially impede retrieval efficiency. Therefore, considering that different factors should contribute differently in the final matching evaluation, we further introduce a final matching token for each query/target, which can be derived by 
adaptively aggregating the corresponding multiple latent factor tokens. Accordingly, this module performs two kinds of aggregations: 1) local feature aggregation for improving the latent factor token representations, and 2) latent factor aggregation for deriving the final matching token representation.

To accomplish the above two kinds of aggregations, we resort to the masked Transformer~\cite{transformer}, 
which can model the diverse interactions among the inputs with flexible attention masks, to jointly fulfil the local feature aggregation and latent factor aggregation. 
Specifically, we feed three types of tokens to the masked Transformer: 1) local feature tokens, 2) latent factor tokens,  and 3) the matching token. The former two types of tokens participate in the local feature aggregation, while the latter two types are involved in the latent factor aggregation.
As the latent factor tokens have been defined in the former module, here we present how to derive the other two types of tokens. As for the local feature tokens, we resort to the output image and text tokens of the second-to-last layer of CLIP, and employ two fully connected layers to further map the image and text tokens to the same latent space of the latent factor tokens, respectively. Let $\mathbf{E}_r \in \mathbb{R}^{L \times D}$ and $\mathbf{E}_t \in \mathbb{R}^{L \times D}$ denote the unified image tokens of the reference image and the target image, respectively, while $\mathbf{E}_m \in \mathbb{R}^{L' \times D}$ denote the unified text tokens of the modification text. $L$ and $L'$ specify the number of tokens for the image and the text, respectively. Pertaining to the matching token used for summarizing all the latent factor tokens, we resort to the fused global representation of the multimodal query ($\mathbf{f}_q$) and the global feature of the target image ($\mathbf{f}_t$). Concretely, the matching token for the query side is defined as $\mathbf{a}_q = \mathbf{f}_q + \mathbf{m}_q$, while that for the target side is defined as $\mathbf{a}_t = \mathbf{f}_t + \mathbf{m}_t$.  $\mathbf{m}_q$ and $ \mathbf{m}_t$ refer to the corresponding identifier tokens, assisting in differentiating the matching token from the latent factor tokens.

Then for local feature aggregation towards improving the latent factor token representations, we link the former two types of tokens with the original \mbox{self-attention} mechanism in Transformer, to comprehensively model their interactions. Notably, the benefit of using self-attention instead of cross-attention is that it allows not only the inter-interactions between the local feature tokens and latent factor tokens but also the intra-interactions among factor tokens. We believe that it is meaningful to model the intra-interactions among factor tokens, as some latent factors could be correlated. For example, the matching factors of ``sleeve length'' and ``material'' are highly correlated. As for the latent factor aggregation, we resort to the \mbox{cross-attention} mechanism, which allows for the information aggregation from the latent factor tokens to the matching token. The cross-attention can be achieved by masking the corresponding attention maps in the Transformer architecture.

Mathematically, the masked Transformer based dual aggregation for the query and target sides can be formulated as follows,
\begin{equation}
\left\{
\begin{aligned}
&\left[\hat{\mathbf{X}}_{q}; \hat{\mathbf{a}}_{q}\right]=\mathcal{F}_{\operatorname{masked\_Transformer}}\left( \left[ \mathbf{E}_{r}; \mathbf{E}_{m}; \mathbf{X}_{q} ; \mathbf{a}_{q} \right]\right), \\
&\left[\hat{\mathbf{X}}_t; \hat{\mathbf{a}}_{t}\right]=\mathcal{F}_{\operatorname{masked\_Transformer}}\left( \left[ \mathbf{E}_{t}; \mathbf{X}_{t} ; \mathbf{a}_{t} \right]\right),
\end{aligned}\label{eq10}
\right.
\end{equation}
where $\hat{\mathbf{X}}_q \in \mathbb{R}^{U \times D}$ and $\hat{\mathbf{X}}_t \in \mathbb{R}^{U \times D}$ denote the enhanced latent factor token representations of the query and target sides, respectively. $\hat{\mathbf{a}}_{q} \in \mathbb{R} ^ {D}$ and $\hat{\mathbf{a}}_{t} \in \mathbb{R} ^ {D}$ refer to the final query matching token and target matching token that have aggregated information from corresponding latent factor tokens, respectively.
It is also worth noting that the parameters of the masked Transformers that process the query side input and target side input are shared. The underlying philosophy is to make the output of both sides still remain in the same semantic space, and hence facilitate the following dual query-target matching modeling.

\subsubsection{Dual Query-Target Matching Modeling}
Having obtained the matching token of both the query and target sides, we can move to the final query-target matching modeling. Although only the final query/target matching tokens will be used in the inference stage to achieve retrieval efficiency, we simultaneously deploy the commonly used batch-based classification loss over both the final matching tokens and the intermediate latent factor tokens for thorough training optimization. 

The core of the \mbox{batch-based} classification loss is to enforce the representations of the multimodal query close to that of the ground-truth target image in a \mbox{mini-batch}, while far from other target images.
For the matching tokens, we directly compute the query-target matching degree based on their representations. As for the latent factor tokens, we can first calculate the factor-specific matching degree based on each query-target matching factor, which can be achieved by computing the matching degree based on the corresponding latent factor tokens. And then the overall query-target matching degree can be derived by summing all the factor-specific matching degrees.

Mathematically, we have the following two loss functions for optimizing the matching tokens and latent factor tokens, respectively,
\begin{equation}
\begin{footnotesize}
\left\{\begin{aligned}
\mathcal{L}_{rank}^{mat} \! &= \! \frac{1}{B} \sum_{i=1}^{B} \! -\log \left\{ \frac{\exp \left\{ \operatorname{s} \left( {\hat{\mathbf{a}}}_{qi} , {\hat{\mathbf{a}}}_{ti} \right)  / \tau\right\}}{ \sum_{j=1}^{B} \exp \left\{ \operatorname{s} \left( {\hat{\mathbf{a}}}_{qi}, {\hat{\mathbf{a}}}_{tj} \right) / \tau \right\}  } \right\}, \\ 
\mathcal{L}_{rank}^{lat} \! &= \! \frac{1}{B} \sum_{i=1}^{B} \! -\log \left\{ \frac{\exp \! \left\{ \! \left\{ \sum_{u=1}^{U} \operatorname{s} \! \left( {\hat{\mathbf{X}}}_{qi} \! \left[u\right], {\hat{\mathbf{X}}}_{ti} \! \left[u\right] \right) \right\}  / \! \tau \!\right\}}{ \sum_{j=1}^{B} \exp \! \left\{ \! \left\{ \sum_{u=1}^{U} \operatorname{s} \! \left( {\hat{\mathbf{X}}}_{qi}\!\left[u\right], {\hat{\mathbf{X}}}_{tj}\!\left[u\right] \right) \right\} \! / \! \tau \right\}} \right\},
\end{aligned}
\right.\label{eq12}
\end{footnotesize}
\end{equation}
where the subscript $i$ denotes the $i$-th triplet sample in the \mbox{mini-batch}, 
$B$ is the batch size, $\operatorname{s}(\cdot, \cdot)$ denotes the cosine similarity function, and $\tau$ is the temperature factor. ${\hat{\mathbf{a}}}_{qi}$ and ${\hat{\mathbf{a}}}_{ti}$ stand for the matching tokens of the multimodal query and the target image, respectively. 
$\hat{\mathbf{X}}_{qi}\left[u\right]$ and ${\hat{\mathbf{X}}}_{ti}\left[u\right]$ denote the $u$-th row of $\hat{\mathbf{X}}_{qi}$ and ${\hat{\mathbf{X}}}_{ti}$, representing the $u$-th latent factor token representation for the query and target, respectively.

Ultimately, we have the following objective function for optimizing our LIMN,

\begin{equation}
\mathbf{\Theta^{*}}=\underset{\mathbf{\Theta}}{\arg \min } \left( \mathcal{L}_{rank}^{mat} + \mathcal{L}_{rank}^{lat} + \mathcal{L}_{L1} \right),\label{eq13}
\end{equation}
where $\Theta$ denotes the \mbox{to-be-learned} parameters in LIMN. Once LIMN is \mbox{well-trained}, all the gallery images can be ranked by their cosine similarities with the input multimodal query, and the CIR task gets solved. 

\subsection{Iterative Dual Self-Training Paradigm}

\begin{figure*}[ht]
\centering
	\includegraphics[width=0.9\linewidth]{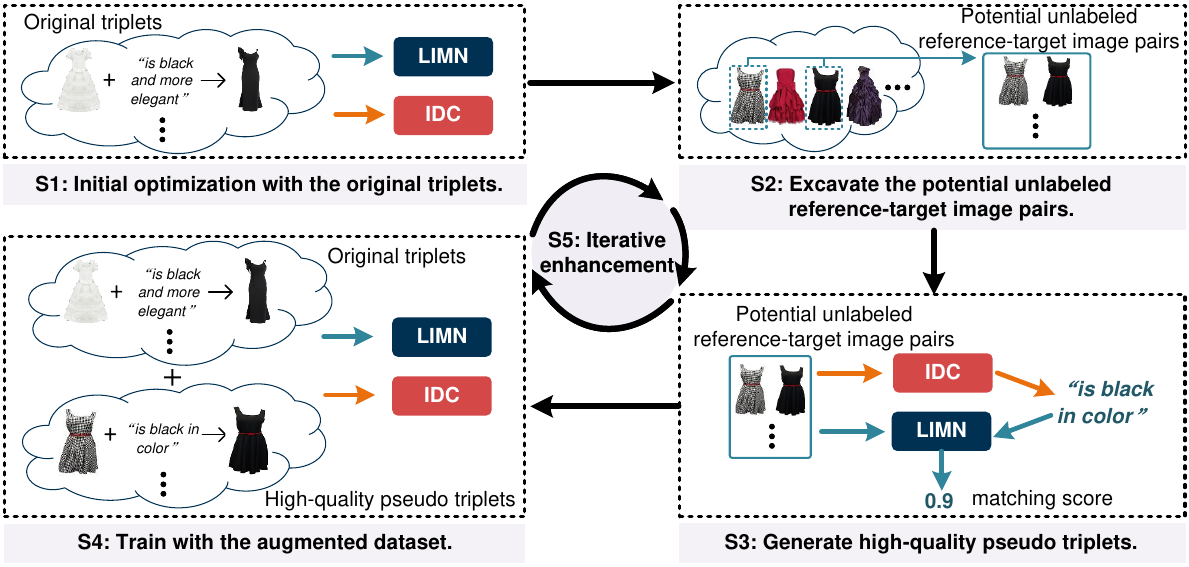}
    \vspace{-0.4em}
	\caption{The proposed iterative dual self-training paradigm for boosting the performance of LIMN and deriving LIMN+, which consists of five steps.}
	\label{fig:self-training}
    \vspace{-1em}
\end{figure*}

Another major novelty of our work is that we resort to the \mbox{self-training} technique to fully exploit the unlabeled data, so as to alleviate the overfitting phenomenon caused by the limited size of training data. Self-training is a commonly used \mbox{weakly-supervised} technique in the image classification field~\cite{self1, self2, self3}, which typically aims to first train a model with the labeled data, and then employ the \mbox{well-trained} model to derive pseudo labeled data by labeling the unlabeled data. Finally, both the labeled and pseudo labeled data are jointly utilized to train another model, which usually yields better performance than the model trained purely with the pre-labeled data.  

To adapt the \mbox{self-training} strategy to the CIR task, we involve a highly correlated task to our \mbox{CIR} task, \textit{i.e.}, image difference captioning (IDC for short)~\cite{duda,acl2018,acl2019,acl2019-2,cvpr21,icmr20}, which aims to generate a natural language-based text to describe the difference between the two relevant images. It is apparent that the dataset for the CIR task can be easily \mbox{re-purposed} for the IDC task, and vice versa. For example, each training triplet for CIR can be re-used for IDC, where the input is the image pair $\left( x_{r}, x_{t} \right)$, while the output is the corresponding modification text $t_{m}$. Therefore, we devise an iterative dual \mbox{self-training} paradigm that takes advantage of the duality between the two highly correlated tasks, to boost the performance in CIR. Specifically, in each iteration, we utilize a \mbox{well-trained} IDC model to generate extra pseudo triplets for augmenting the training dataset of the  CIR model and hence  promoting its generalization ability. Meanwhile, to get rid of the noisy triplets, we employ the \mbox{well-trained} CIR model to evaluate the quality of the pseudo triplets generated by the IDC model. 

Combing our context, we illustrate the devised iterative dual self-training paradigm in Figure~\ref{fig:self-training}, where the CIR model is instantiated with the proposed LIMN, while the IDC model not specified as any advanced IDC model is applicable. Notably, our proposed iterative dual self-training paradigm can be also applied to other CIR models to boost their performance. As can be seen from Figure~\ref{fig:self-training}, the iterative dual self-training paradigm can be summarized with five steps as follows. 

\begin{itemize}
    \item[\textbf{S1:}] \textbf{Initial optimization with the original triplets.} 
    Optimize the LIMN model and the IDC model based on the original training dataset, respectively. Let $\Theta_{*}^{'}$ denote the parameters of the LIMN model and $\Phi_{*}^{'}$ indicate the parameters of the IDC model.
    \item[\textbf{S2:}] \textbf{Excavate the potential unlabeled \mbox{reference-target} image pairs.} 
    Following the original \mbox{reference-target} image pair construction method of each dataset, we can excavate all the potential unlabeled \mbox{reference-target} image pairs from the training dataset $\mathcal{T}$,
    denoted as $\left\{ \left( \hat{x}_{r}, \hat{x}_{t} \right)_{j} \right\}_{j=1}^{M}$. Note that different datasets differ in constructing image pairs, which will be detailed in Section $4$.
    \item[\textbf{S3:}] \textbf{Generate high-quality pseudo triplets.} Employ the well-trained IDC model to generate the modification text that describes the difference between the potential unlabeled \mbox{reference-target} image pairs, which is \mbox{as follows},
    \begin{equation}
        \tilde{t}_{m} = \operatorname{IDC}\left(\left(\hat{x}_r, \hat{x}_t\right), \Phi_{*}^{'}\right).
        \label{eq14}
    \end{equation}
    In this way, we can obtain the pseudo triplets $\left( \hat{x}_{r}, \tilde{t}_{m} ,\hat{x}_{t} \right)$.
    To ensure the quality of pseudo triplets, we utilize the \mbox{well-trained} LIMN model to measure the matching score between the multimodal query and the target image in each pseudo triplet, \textit{i.e.},
    \begin{equation}
        s = \operatorname{LIMN}\left(\left(\hat{x}_r, \tilde{t}_m, \hat{x}_t\right), \Phi_{*}^{'}\right).
        \label{eq15}
    \end{equation}
    Thereafter, the pseudo triplets are sorted in descending order according to their corresponding matching scores $s$. The top $\kappa$ pseudo triplets with high quality will be retained, and others will be removed.
    \item[\textbf{S4:}] \textbf{Train with the augmented dataset.} The high-quality pseudo triplets and the original triplets are alternatively used to re-train the LIMN model $\Theta_{*}^{''}$ and the IDC model $\Phi_{*}^{''}$, respectively. 
    \item[\textbf{S5:}] \textbf{Iterative enhancement.} Repeat step2 - step4 until the performance of LIMN no longer increases, and it can be exercised during the inference stage.
\end{itemize}
For clarity, we denote the iterative dual \mbox{self-training} paradigm boosted LIMN as LIMN+.

Note that in \textbf{S4}, we empirically observed that directly mixing up the original triplets with the pseudo triplets can negatively affect the CIR performance. This may be because although we have utilized the well-trained LIMN model to filter the high-quality pseudo data, the pseudo triplets are still not as reliable as the original manually labeled ones. 
To address this, we design an asymmetric alternating training strategy. Specifically, the original triplets and pseudo triplets are used alternatively for training with an alternative ratio $n$. This ratio specifies that the model would be first trained with the original triplets for $n$ epochs, then with the pseudo triplets for one epoch. This can alleviate the negative impact brought
by the lower quality of the pseudo dataset.

\section{Experiment}
In this section, we first give the experimental settings and then detail the experiments conducted on four datasets by answering the following research  questions:
\begin{itemize}
    \item \textbf{RQ1}: Do our models (LIMN and LIMN+)  surpass \mbox{state-of-the-art} methods?
    \item \textbf{RQ2}: How does each component affect LIMN and LIMN+?
    \item \textbf{RQ3}: Does the devised iterative dual self-training paradigm boost the model performance?
    \item \textbf{RQ4}: \mbox{How is the quantitative performance of LIMN+?}
\end{itemize}

\subsection{Experimental Settings}
\subsubsection{Datasets}  
There have been several public datasets in the domain of CIR. 
According to the modification text construction manner, the datasets can be classified into two groups: \mbox{attribute-based} datasets~\cite{fashion200k,mitstates,tirg} and natural \mbox{language-based} datasets~\cite{fashioniq,shoes,cirr}. The modification text in the former group is synthesized by \mbox{pre-specified} templates, such as ``\textit{replace A with B}'', which are not flexible and the application prospect is limited. In contrast, the modification text of datasets in the latter group is \mbox{human-annotated}, which is more flexible for users to express the modification demands.
Toward comprehensive evaluation, we chose four datasets: FashionIQ~\cite{fashioniq}, Shoes~\cite{shoesguo}, CIRR~\cite{cirr}, and Fashion200K~\cite{fashion200k}. The former three are \mbox{real-world} natural \mbox{language-based} datasets, which can be used for evaluating both LIMN and LIMN+, while the last one is an \mbox{attribute-based} dataset which is constructed automatically without the scarce data problem and can be used only for evaluating LIMN. 

\textbf{FashionIQ~\cite{fashioniq}} is an interactive fashion retrieval dataset. The fashion items within the dataset belong to three categories: Dresses, Tops\&Tees, and Shirts. In our experiments, we treated them as three separate datasets.
Following~\cite{val, dcnet, clvcnet}, \mbox{$\sim$\hspace{0em}$46$K} images and \mbox{$\sim$\hspace{0em}$15$K} images are used for training and testing, respectively. Finally, there are $18$K triplets for training and \mbox{$\sim$\hspace{0em}$6$K} for testing.

\textbf{Shoes~\cite{shoesguo}} is originally created for the attribute discovery task by~\cite{shoes} and is then developed with relative caption annotations for the dialog-based interactive retrieval task~\cite{shoesguo}.
Following~\cite{val, dcnet, clvcnet}, $10$K images are used for training and \mbox{$\sim$\hspace{0em}$4.6$K} images for testing, in which \mbox{$\sim$\hspace{0em}$9$K} triplets and \mbox{$\sim$\hspace{0em}$1.7$K} triplets are constructed, respectively.

\textbf{CIRR}~\cite{cirr} is an open-domain dataset, comprising \mbox{$\sim$\hspace{0em}$21$K} real-life images taken from the NLVR$^{2}$ dataset~\cite{nlvr}. It contains \mbox{$\sim$\hspace{0em}$36.5$K} triplets in total, where $80$\% for training, $10$\% for validation, and $10$\% for testing.

\textbf{Fashion200K}~\cite{FCD} contains \mbox{$\sim$\hspace{0em}$200$K} fashion images, each with an \mbox{attribute-like} text description. Following~\cite{tirg,val}, the pair of images are constructed by identifying only \mbox{one-word} difference in their descriptions, while the modification text is synthesized by templates, such as ``replace \textit{red} with \textit{green}''. There are \mbox{$\sim$\hspace{0em}$172$K} triplets for training and \mbox{$\sim$\hspace{0em}$33$K} triplets for evaluation. 

\subsubsection{Implementation Details} 
LIMN utilizes pre-trained CLIP~\cite{clip}(ViT-L/14 version) as the backbone. The feature dimension $D$ is set to $768$. The \mbox{parameter-shared} masked Transformer for processing the multimodal query and the target image is set to $2$ layers and $8$ heads. The temperature factor in Eqn.~(\ref{eq12}) is set to $10$. We trained LIMN by AdamW optimizer with an initial learning rate of $1$e-$4$. Specifically, the learning rate for CLIP is set to $1$e-$5$ for better convergence. The learning rate decays by a factor of $0.1$ at the $5$-th epoch. We empirically set the batch size as $32$. The number of matching factors $U$ is set as $8$, $8$, $10$, and $8$ for FashionIQ, Shoes, CIRR, and Fashion200K through hyperparameter tuning, respectively. For a fair comparison, we utilized the same evaluation metric, \textit{i.e.}, recall at rank $k$ (R@$k$), as previous efforts~\cite{val,dcnet,clvcnet,clip4cir,cirr} to measure the image retrieval performance. 

Pertaining to the iterative dual self-training paradigm, we directly utilized the \mbox{off-the-shelf} DUDA~\cite{duda} as the IDC model~\footnote{Any other IDC model is acceptable.}. We trained DUDA by Adam optimizer with a fixed learning rate of $1$e-$4$, where the batch size is set to $32$. We resorted to the \mbox{widely-used} \mbox{BLEU-$1$}~\cite{bleu} and \mbox{ROUGE-L}~\cite{rouge} metrics to evaluate the quality of the generated pseudo triplets. The numbers of the retained pseudo triplets $\kappa$ are empirically set to \mbox{$6$,$000$} for each category in FashionIQ, \mbox{$9$,$000$} for Shoes, and \mbox{$10$,$000$} for CIRR, respectively, which are comparable to the size of the original training set of the three datasets. The alternating training ratio $n$ is set through grid search among $\left[2,3,4 \right]$ for three datasets.

\subsubsection{Potential Unlabeled Reference-Target Image Pairs Construction}
According to the Step2 of our iterative dual \mbox{self-training} paradigm, we need to find the potential unlabeled \mbox{reference-target} image pairs for each dataset. Notably, as aforementioned, only the natural language-based datasets FshionIQ, Shoes, and CIRR are suitable for evaluating our iterative dual self-training paradigm. Therefore, here we present how to construct the potential unlabeled reference-target image pairs for these three datasets. To ensure the distribution of pseudo triplets as consistent as possible with that of each original dataset, we attempted to excavate the potential unlabeled \mbox{reference-target} image pairs in the same manner as the original image pair construction of each dataset \mbox{as follows}.

\begin{table*}[t!]
    \centering
    \caption{Performance comparison on FashionIQ and Shoes with respect to R@$k$($\%$). The best results over baselines are underlined, while the overall best results are in boldface. The missing results of some methods are because they did not report their results on the Shoes dataset. The penultimate row quantifies the absolute performance improvements of LIMN+ over LIMN. The last row indicates the performance improvements by LIMN+ over the best baseline. The results of the proposed method are marked with a gray background.
    }\label{tab:exp_fashioniq_shoes}
    \begin{tabular}{l|cc|cc|cc|cc||ccc}
    \hline 
    \multirow{3}{*}{Method} &\multicolumn{8}{c||}{FashionIQ} & \multicolumn{3}{c}{\multirow{2}{*}{Shoes}} \\ 
    \cline{2-9} 
    & \multicolumn{2}{c|}{Dresses} & \multicolumn{2}{c|}{Shirts} & \multicolumn{2}{c|} {Tops\&Tees} & \multicolumn{2}{c||}{Avg} & & & \\ \cline{2-12}  & R@$10$ & R@$50$ & R@$10$ & R@$50$ & R@$10$ & R@$50$ & R@$10$ & R@$50$ & R@$10$ & R@$50$ & Avg \\
    \hline \hline 

    TIRG~\cite{tirg} & $14.87$ & $34.66$ & $18.26$ & $37.89$ & $19.08$ & $39.62$ & $17.40$ & $37.39$ & $45.45$ & $69.39$ & $57.42$ \\
    VAL~\cite{val}  & $21.12$ & $42.19$ & $21.03$ & $43.44$ & $25.64$ & $49.49$ & $22.60$ & $45.04$ & $49.12$ & $73.53$ & $61.33$ \\
    CIRPLANT~\cite{cirr} & $17.45$ & $40.41$ & $17.53$ & $38.81$ & $21.64$ & $45.38$ & $18.87$ & $41.53$ & $-$ & $-$ \\
    CosMo~\cite{cosmo}  & $25.64$ & $50.30$ & $24.90$ & $49.18$ & $29.21$ & $57.46$ & $26.58$ & $52.31$ & $48.36$ & $75.64$ & $62.00$ \\
    DCNet~\cite{dcnet} & $28.95$ & $56.07$ & $23.95$ & $47.30$ & $30.44$ & $58.29$ & $27.78$ & $53.89$ & $53.82$ & $79.33$ & $66.58$ \\
    DATIR~\cite{datir}  & $21.90$ & $43.80$ & $21.90$ & $43.70$ & $27.20$ & $51.60$ & $23.70$ & $46.40$ & $51.10$ & $75.60$ & $63.35$ \\
    CLVC-Net~\cite{clvcnet} & $29.85$ & $56.47$ &$ 28.75$ & $54.76$ & $33.50$ & $64.00 $& $30.70$ &$ 58.41 $  & $54.39$ & $79.47$ & $66.93$ \\
    ARTEMIS~\cite{artemis}  & $27.16$ & $52.40$ & $21.78$ & $43.64$ &$ 29.20 $& $54.83 $& $26.05 $& $50.29 $  & $53.11$ & $79.31$ & $66.21$ \\
    SAC~\cite{SAC} & $26.52$ & $51.01$ & $28.02$ & $51.86$ & $32.70$ & $61.23$ & $29.08$ & $54.70$ & $51.73$ & $77.28$ & $64.51$ \\
    EER~\cite{tip22} & $30.02$ & $55.44$ & $25.32$ & $49.87$ & $33.20$ & $60.34$ & $29.51$ & $55.22$  & $56.02$ & $79.94$ & $67.98$\\
    CRR~\cite{crr} & $30.41$ & $57.11$ & $30.73$ & $58.02$ & $33.67$ & $64.48$ & $31.60$ & $59.87$ & $56.38$ & $79.92$ & $68.15$\\
    FashionVLP~\cite{fashionvlp}& $32.42$ & $60.29$ & $31.89$ & $58.44$ & $38.51$ & $68.79$ & $34.27$ & $62.51$ & $49.08$ & $77.32$ & $63.20$ \\
    ComqueryFormer~\cite{comqueryformer} & $33.86$ & $61.08$ & $35.57$ & $62.19$ & $42.07$ & $69.30$ & $37.17$ & $64.19$ & $-$ & $-$  & $-$ \\
    AMC~\cite{amc} & $31.73$ & $59.25$ & $30.67$ & $59.08$ & $36.21$ & $66.60$ & $32.87$ & $61.64$ & $56.89$ & $79.27$ & $68.08$ \\
    \hdashline
    Clip4cir~\cite{clip4cir} & $33.81$ & $59.40$ & $39.99$ & $60.45$ & $41.41$ & $65.37$ & $38.32$ & $61.74$ & $-$ & $-$ & $-$ \\
    Prog. Lrn.~\cite{zhao}& \underline{$38.18$} & \underline{$64.50$} &\underline{$48.63$}& \underline{$71.54$} & \underline{$52.32$} & \underline{$76.90$} & \underline{$46.37$} &\underline{$70.98$} & \underline{$58.83$} & \underline{$84.16$} & \underline{$71.50$} \\
    \hline \hline
   \rowcolor{gray!25} LIMN & $50.72$ & $74.52$ & $56.08$ & $77.09$ & $60.94$ & $81.85$ & $55.91$ & $77.82$ & $68.20$ & $87.45$ & $77.83$ \\
   \rowcolor{gray!32} \textbf{LIMN+} & $\mathbf{52.11}$
    & $\mathbf{75.21}$ & $\mathbf{57.51}$ & $\mathbf{77.92}$ & $\mathbf{62.67}$ & $\mathbf{82.66}$ & $\mathbf{57.43}$ & $\mathbf{78.60}$ & $\mathbf{68.37}$ & $\mathbf{88.07}$ & $\mathbf{78.22}$  \\
    \hdashline
    $\Delta$ LIMN+ & $+1.39$ & $+0.69$ & $+1.43$ & $+0.83$ & $+1.73$ & $+0.81$ & $+1.52$ & $+0.78$ & $+0.17$ & $+0.62$ & $+0.39$ \\
     Improvement(\%) & $\uparrow 36.49 $ & $\uparrow 16.60 $ & $\uparrow 18.26 $ & $\uparrow 8.92 $ & $\uparrow 19.78 $ & $\uparrow 7.49 $ & $\uparrow 23.85 $ & $\uparrow 10.74 $ & $\uparrow 16.22$ & $\uparrow 4.65$ & $ \uparrow 9.40 $ \\
    \hline
    \end{tabular}
    \vspace{-0.5em}
\end{table*}

\begin{table*}[t!]
    \centering \caption{Performance comparison on CIRR with respect to R@$k$($\%$) and R$_{subset}$@$k$($\%$). The best baseline results are underlined, while the overall best results are in boldface. The penultimate row quantifies the absolute performance improvements of LIMN+ over LIMN. The last row indicates the performance improvements by LIMN+ over the best baseline. The results of the proposed method are marked with a gray background.
    }\label{tab:exp_cirr}

    \begin{tabular}{l|cccc|ccc|c}
    \hline 
    \multirow{2}{*}{Method} &\multicolumn{4}{c|}{\textbf{R@$k$}} &\multicolumn{3}{c|}{\textbf{R$_{subset}$@$k$}} & \multirow{2}{*}{(R@$5$ + R$_{subset}$@$1$) / $2$} \\ \cline{2-8}
    & $k=1$ & $k=5$ & $k=10$ & $k=50$ & $k=1$ & $k=2$ & $k=3$ \\
    \hline \hline 
    TIRG~\cite{tirg} & $14.61$ & $48.37$ & $64.08$ & $90.03$ & $22.67$ & $44.97$ & $65.14$ & $35.52$\\
    ARTEMIS~\cite{artemis} & $16.96$ & $46.10$ & $61.31$ & $87.73$ &$ 39.99 $& $62.20 $& $75.67 $ & $43.05$\\
    CIRPLANT~\cite{cirr}  & $15.18$ & $43.36$ & $60.48$ & $87.64$ & $33.81$ & $56.99$ & $75.40$ & $38.59$\\
    ComqueryFormer~\cite{comqueryformer} & $25.76$ & $61.76$ & $75.90$ & $95.13$ & $51.86$ & $76.26$ & $89.25$ & $56.81$\\
    \hdashline
    Clip4cir~\cite{clip4cir}  & \underline{$38.53$} & \underline{$69.98$} &\underline{$81.86$}& \underline{$95.93$} & \underline{$68.19$} & \underline{$85.64$} & \underline{$94.17$} & \underline{$69.09$}\\
    \hline \hline
   \rowcolor{gray!25} LIMN & $\mathbf{43.64}$ & $75.37$ & $85.42$ & $97.04$ & $69.01$ & $86.22$ & $94.19$ & $72.19$ \\
   \rowcolor{gray!32} \textbf{LIMN+} & $43.33$ & $\mathbf{ 75.41}$ & $\mathbf{85.81}$ & $\mathbf{97.21}$ & $\mathbf{ 69.28}$ & $\mathbf{ 86.43}$ & $\mathbf{94.26 }$ & $\mathbf{ 72.35}$ \\
    \hdashline
    $\Delta$ LIMN+ & $-0.31$ & $+0.04$ & $+0.39$ & $+0.17$ & $+0.27$ & $+0.21$ & $+0.07$ & $ +0.16$ \\ 
    Improvement(\%) & $\uparrow 12.46 $ & $\uparrow 7.76 $ & $\uparrow 4.83 $ & $\uparrow 1.33 $ & $\uparrow 1.60 $ & $ \uparrow 0.92 $ & $ \uparrow 0.10 $ & $ \uparrow 4.72 $ \\ 
    \hline
    \end{tabular}
    \vspace{-0.3em}

\end{table*}

1) For the {FashionIQ} dataset, following~\cite{fashioniq}, for each training image, we first computed the \mbox{TF-IDF} score of each word in its corresponding fashion item title. We then paired each training image with another training image that achieves the maximum value of the summation over the \mbox{TF-IDF} scores on all overlapping words between the two items' titles. Finally, we obtained \mbox{$\sim$\hspace{0em}$30$K} potential unlabeled \mbox{reference-target} image pairs for the FashionIQ dataset, of which \mbox{$7$,$741$} belong to the category of Dress, \mbox{$9$,$925$} to the Tops\&Tees, and \mbox{$12$,$062$} to the Shirts. 

2) Regarding the Shoes dataset, as~\cite{shoesguo} did not release the details for constructing the \mbox{reference-target} image pairs, we proposed a \mbox{statistic-based} strategy to find the potential unlabeled \mbox{reference-target} image pairs. In particular, we first computed the similarity score of each \mbox{reference-target} image pair in the training set with our  \mbox{pre-trained} LIMN. It shows that the mean and variance of the similarity scores in the whole training set are $0.9703$ and $0.0153$, respectively. Accordingly, we randomly selected two images, and retained them as long as their similarity score falls in the range of $\left[0.9703-0.0153, 0.9703+0.0153\right]$. Ultimately, we obtained \mbox{$20$,$000$} potential unlabeled \mbox{reference-target} image pairs. 

3) As for the CIRR dataset, the authors~\cite{cirr} extracted a large collection of image pairs from the NLVR$^2$ dataset using specific criteria, and then selected a random subset for CIRR dataset. However, when we attempted to construct unlabeled reference-target image pairs using the same approach, challenges arise. We were restricted to using the CIRR training set to avoid external data. Since this set is a curated subset from NLVR$^2$, applying the same criteria could result in significant overlaps and repetition of image pairs already in the CIRR dataset. 
To address this, similar to the approach for the Shoes dataset, we calculated the mean and variance of the similarity scores for the \mbox{reference-target} image pairs within the CIRR training dataset, which are $0.6721$ and $0.1816$, respectively. Image pairs were then selected based on the similarity score range of $\left[0.6721-0.1816, 0.6721+0.1816\right]$. This process resulted in a collection of \mbox{$33$,$800$} potential unlabeled \mbox{reference-target} image pairs.

\subsection{Performance Comparison (RQ1)}
To justify the effectiveness of our proposal, we chose the following baselines for comparison: TIRG~\cite{tirg}, VAL~\cite{val}, CIRPLANT\cite{cirr}, CosMo~\cite{cosmo}, DCNet~\cite{dcnet}, DATIR~\cite{datir}, JGAN~\cite{JGAN}, CLVC-Net~\cite{clvcnet}, ARTEMIS~\cite{artemis}, SAC~\cite{SAC}, EER~\cite{tip22}, CRR~\cite{crr}, FashionVLP~\cite{fashionvlp}, ComqueryFormer~\cite{comqueryformer}, AMC~\cite{amc}, Clip4cir~\cite{clip4cir}, and Prog. Lrn.~\cite{zhao}.
Notably, the former fifteen baselines use traditional models like ResNet~\cite{resnet} and LSTM~\cite{lstm} as the feature extraction backbone, while the latter two take advantage of the multimodal pre-trained large model CLIP.
Table~\ref{tab:exp_fashioniq_shoes}, Table~\ref{tab:exp_cirr}, and Table~\ref{tab:fashion200k_exp} summarize the performance comparison among different methods on the four datasets. From these tables, we have the following observations. 
1) Our LIMN+ model consistently outperforms all the baseline methods over the three natural \mbox{language-based} datasets FashionIQ, Shoes, and CIRR by a considerable margin. Specifically, LIMN+ gains $22.61\%$ and $20.18\%$ improvements on FashionIQ-Avg and Shoes over the best baseline with respect to R@$10$, respectively, and $ 12.46\%$ improvement on CIRR regarding R@$1$.
This confirms the superiority of our LIMN+ that utilizes the latent factor tokens to capture the multiple \mbox{query-target} matching factors, as well as employs the iterative dual \mbox{self-training} paradigm to relieve the overfitting problem caused by the limited dataset size. 
2) Without the self-training paradigm, our LIMN model also consistently surpasses all the baseline methods on the four datasets. This observation confirms the powerful capability of our LIMN for the CIR task. 
3) LIMN+ consistently improves the performance of LIMN on the three natural language-based datasets with the iterative dual self-training paradigm. Moreover, the improvements on the FashionIQ dataset are the most prominent, while those on CIRR are the least. This could be attributed to that the scale of the FashionIQ dataset is the smallest with approximately $6,000$ triplets for each category, making the scarce dataset problem of FashionIQ dataset more pronounced. Therefore, augmenting the dataset with self-training can greatly improve the model performance. In contrast,  CIRR, being the largest dataset with nearly $28,000$ training triplets, already has abundant training samples. Therefore, the benefit of using the iterative dual self-training paradigm is limited.

\begin{table}[!t]
	\centering \caption{Performance comparison on Fashion200K. The best baseline results are underlined, while the overall best results are in boldface. The last row indicates the performance improvements by LIMN over the best baseline.}\label{tab:fashion200k_exp}
 \resizebox{0.85\linewidth}{!}{
	\begin{tabular}{l|ccc}
		\hline Method & R@$10$ & R@$50$ & Avg \\
		\hline \hline
		TIRG~\cite{tirg} & $42.5$ & $63.8$ & $53.2$  \\
        VAL~\cite{val} & $49.0$ & $68.8$ & $58.9$ \\	
        DATIR~\cite{datir} & $48.8$ & $71.6$ & $60.2$ \\	
        JGAN~\cite{JGAN} & $45.3$ & $65.7$ & $55.5$  \\
        CosMo~\cite{cosmo} & $50.4$ & $69.3$ & $59.9$ \\	
        DCNet~\cite{dcnet} & $46.9$ & $67.6$ & $57.3$  \\
        CLVC-Net~\cite{clvcnet} & $53.0$ & $72.2$ & $62.6$ \\	
        ARTEMIS~\cite{artemis} & $51.1$ & $70.5$ & $60.8$ \\	
        CRR~\cite{crr} & \underline{$56.4$} & \underline{$73.6$} & \underline{$65.0$} \\
	FashionVLP~\cite{fashionvlp} & $49.9$ & $70.5$ & $60.2$ \\	
        ComqueryFormer~\cite{comqueryformer} & $52.2$ & $72.2$ & $62.2$ \\	
	\hline \hline
	\textbf{LIMN} &$\mathbf{57.2}$ & $\mathbf{76.6}$ & $\mathbf{66.9}$ \\
	\hdashline
      Improvement(\%) & $\uparrow 1.4 $ & $\uparrow 4.1 $ & $\uparrow 2.9 $  \\ 
    \hline
	\end{tabular}}
\end{table}

\begin{table}[t!]
    \centering \caption{Ablation study on FashionIQ, Shoes, and CIRR. 
    }\label{exp:ablation}
    \resizebox{0.95\linewidth}{!}{
    \begin{tabular}{l|c|c|c|c}
    \hline 
    \multirow{2}{*}{Method} &\multicolumn{2}{c|}{FashionIQ-Avg} & Shoes & CIRR \\ \cline{2-5}
    & R@$10$ & R@$50$ & Avg  & Avg  \\
    \hline \hline 
    w/ OneFactor & $ 51.77 $ & $ 74.64 $ & $ 76.32 $ & $ 68.70$ \\
    w/o Factor\_BBC & $ 54.43 $ & $ 76.68 $ & $ 77.09 $ & $ 69.80$ \\
    w/o Global\_Feature & $ 43.92 $ & $ 68.97 $ & $ 71.16 $ & $ 60.91$ \\
    \textbf{LIMN} & $ \mathbf{55.91} $ & $ \mathbf{77.82} $ & $ \mathbf{77.83} $ & $ \mathbf{72.19}$ \\
    \hline \hline 
    w/o Filter & $ 57.17 $ & $ 78.28 $ & $ 77.91 $ & $ 72.03 $ \\
    \textbf{LIMN+} & $ \mathbf{57.43} $ & $ \mathbf{78.60} $ & $ \mathbf{78.22} $ & $ \mathbf{72.35} $ \\

    \hline
    \end{tabular}}
\end{table}

\subsection{Ablation Study (RQ2)}
In this section, we conducted in-depth analyses of each component in LIMN and LIMN+ to explore their effectiveness.
\subsubsection{Influence of Key Components}
To verify the importance of each key component in our method, we compared LIMN and LIMN+ with their following derivatives.

\begin{itemize}
    \item \textbf{w/ OneFactor:} To investigate the effect of introducing multiple latent factor tokens, we used only one latent factor token by setting $U=1$.
    \item \textbf{w/o Factor\_BBC:} To study the effect of explicitly constraining the intermediate latent factor tokens with the batch-based classification loss, we removed the $\mathcal{L}_{rank}^{lat}$ in Eqn.~(\ref{eq13}).
    \item \textbf{w/o Global\_Feature:} To explore the impact of utilizing global features to derive the latent factor tokens and matching tokens, we removed the global features and directly used the identifier tokens by setting $\mathbf{X}_{q} = \left[ \mathbf{m}_{q}^{1},\dots, \mathbf{m}_{q}^{U} \right]$, $\mathbf{X}_{t} = \left[ \mathbf{m}_{t}^{1},\dots, \mathbf{m}_{t}^{U} \right]$, $\mathbf{a}_{q}=\mathbf{m}_{q}$, and $\mathbf{a}_{t}=\mathbf{m}_{t}$.
    \item \textbf{w/o Filter:} To check the importance of filtering high-quality pseudo triplets in the iterative dual self-training paradigm, we removed this filter and directly sampled an equivalent number of pseudo triplets for dataset augmentation.
\end{itemize}

Table~\ref{exp:ablation} shows the ablation results of LIMN on FashionIQ, Shoes, and CIRR datasets. From this table, we gained the following observations. 
1) \mbox{w/ OneFactor} performs worse than our LIMN, which proves the necessity of modeling multiple latent matching factors.  
2) \mbox{w/o Factor\_BBC} also shows inferior results compared to LIMN. This indicates that there remains a tangible benefit in explicitly optimizing these latent factor tokens for deriving a better matching token and hence obtaining superior retrieval performance.
3) \mbox{w/o Global\_Feature} significantly underperforms LIMN. This shows the value of inheriting the rich information from the CLIP-based global features to learn representations for the latent factor tokens and the matching token. Notably, in the absence of these global features, the latent factor tokens can be seen as randomly initialized and remain the same across different samples~\cite{token1,token2}. Therefore, this result also reflects the superiority of our design over conventional schemes.
4) \mbox{w/o Filter} performs worse than LIMN+. This proves the effectiveness of utilizing the well-trained CIR model to filter the pseudo samples generated by the IDC model and obtain high-quality ones.

\begin{figure*}[t!]
    \centering
	\includegraphics[width=0.95\linewidth]{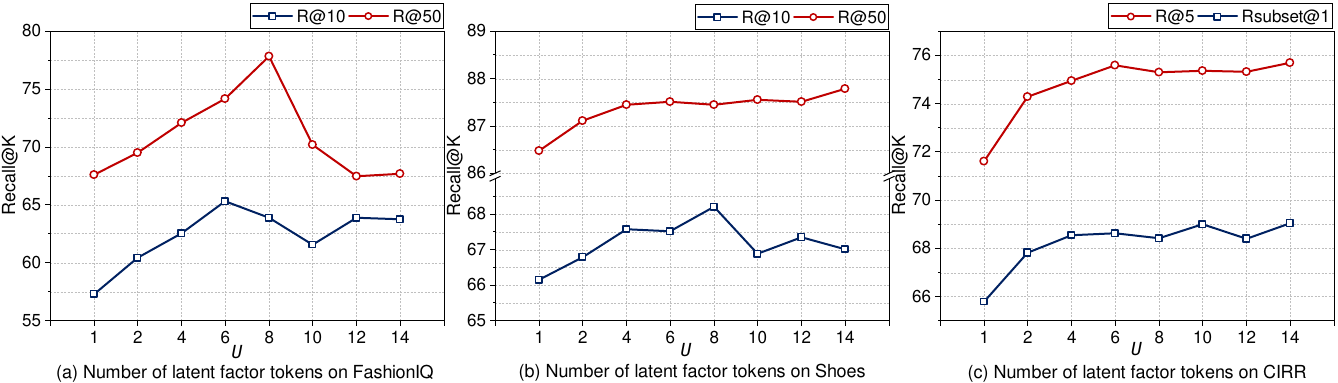}
    \vspace{-1em}
	\caption{Influence of the number of latent factor tokens $U$ on (a) FashionIQ, (b) Shoes, and (c) CIRR.}\label{fig:template}
\end{figure*}

\subsubsection{Number of Latent Factor Tokens}
To provide more insight into the effect of multiple latent factor tokens in LIMN, we illustrated the performance of our LIMN with respect to different numbers of latent factor tokens on FashionIQ, Shoes, and CIRR in Figure~\ref{fig:template}. Notably, the range of the number of latent factor tokens explored for each dataset is set according to the corresponding optimal value of $U$. From Figure~\ref{fig:template}, we observed that as the number of latent factor tokens increases, the performance of LIMN on three datasets first rises and then either decreases (Figure~\ref{fig:template}(a)) or tends to be stable (Figure~\ref{fig:template}(b)(c)). This is reasonable since more latent factor tokens can cover more query-target matching factors, and the retrieval performance will be better. Nevertheless, too many latent factor tokens will not bring additional performance improvement. This may be due to the fact that the number of useful query-target matching factors in real-world application scenarios is limited and thus introducing more latent factor tokens is redundant and can lead to overfitting.
Empirically, we propose that using $8$ latent factor tokens serves as a reliable baseline and a practical starting point for hyper-parameter tuning in real-world application scenarios.

\begin{figure*}[ht]
    \centering
	\includegraphics[scale=1.25]{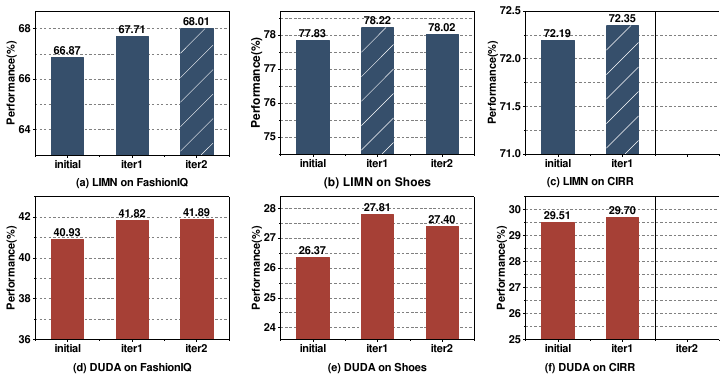}
 \vspace{-1em}
	\caption{
  The iterative performance of our CIR model LIMN and the IDC model DUDA under the iterative dual \mbox{self-training} paradigm on FashionIQ, Shoes, and CIRR datasets. The performance of CIR is the average of R@$k$, and that of IDC is the average between \mbox{BLEU-$1$} and \mbox{ROUGE-L}.
  The initial bar in each plot refers to the initial performance of the original model, while the iter$*$ bar denotes the $*$-th self-training iteration, respectively. The iteration stops when the performance gains are very limited (a and c) or the performance decreases (b).
  The bars with white slashes indicate the optimal performance reported in this work.}\label{fig:iteration}
  \vspace{-1em}
\end{figure*}

\subsection{Iterative Dual Self-training Paradigm (RQ3)}
In this subsection, to thoroughly explore the effectiveness of the proposed iterative dual \mbox{self-training} paradigm, we conducted experiments to investigate the \mbox{self-training} iterative performance of the CIR and IDC models, as well as the generalization ability of the iterative dual \mbox{self-training} paradigm towards CIR. 

\subsubsection{Self-training Iterative Performance}
To gain a more comprehensive understanding of our proposed iterative dual \mbox{self-training} paradigm, we investigated the performance of our LIMN model and the IDC model, \textit{i.e.}, DUDA, in each iteration under the iterative dual self-training paradigm on three datasets. As illustrated in Figure~\ref{fig:iteration}, compared to the initial performance, the iterative dual \mbox{self-training} paradigm consistently boosts the image retrieval performance of LIMN and the image difference captioning performance of DUDA on three datasets. This confirms that the advantage of  our proposed iterative dual \mbox{self-training} paradigm in fully utilizing the potential unlabeled \mbox{reference-target} image pairs and relieving the setback of the limited training data. Moreover, as can be seen, in most cases, the performance improvement by only one \mbox{self-training} iteration is significant, which suggests the high efficiency of our iterative dual self-training paradigm.

\begin{table}[ht]
    \centering \caption{Self-training boosted CLVC-Net performance. The results on FashionIQ are the average result of three categories.}\label{tab:baseline+}
    \vspace{-0.7em}
    \resizebox{0.96\linewidth}{!}{
    \begin{tabular}{l|cc||cc}
    \hline 
    \multirow{2}{*}{Method} & \multicolumn{2}{c||}{FashionIQ-Avg} & \multicolumn{2}{c}{Shoes} \\
    \cline{2-5} & R@$10$ & R@$50$ & R@$10$ & R@$50$  \\
    \hline \hline 
   CLVC-Net  & $30.70$ & $58.41$ & $54.39$ & $79.47$  \\
   CLVC-Net+  & $31.55$ & $60.24$ & $55.07$ & $80.12$  \\
   \hline
   Improvement(\%) & $\uparrow 2.77$ & $\uparrow 3.13$ & $\uparrow 1.25$ & $\uparrow 0.82$   \\
    \hline
    \end{tabular}}
    \vspace{-1em}
\end{table}

\begin{figure*}[t]
\centering
	\includegraphics[width=0.98\linewidth]{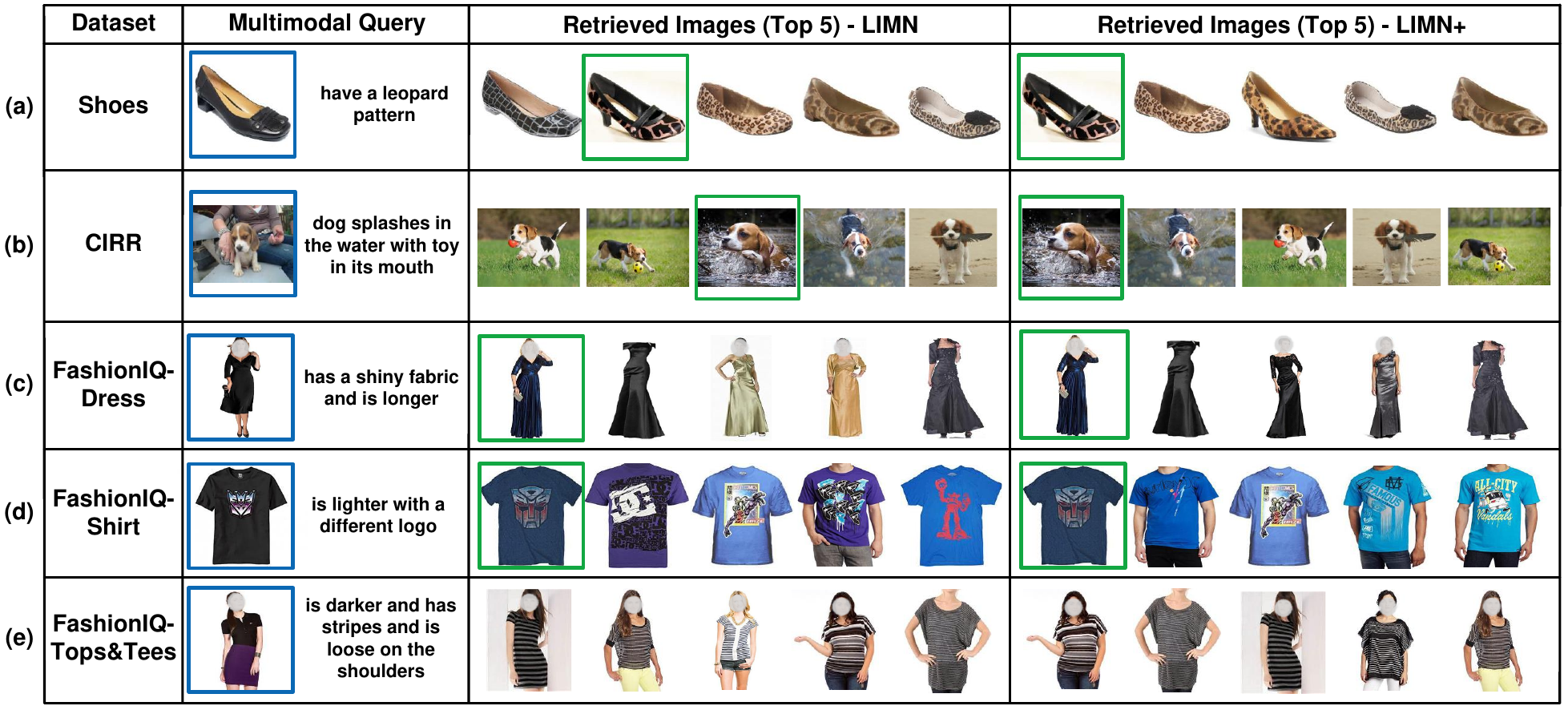}
    \vspace{-0.5em}
	\caption{Illustration of several CIR results obtained by our LIMN and LIMN+ on three datasets.}\label{fig:case_study}
\end{figure*}

\begin{figure}[!t]
	\centering
	\includegraphics[width=\linewidth]{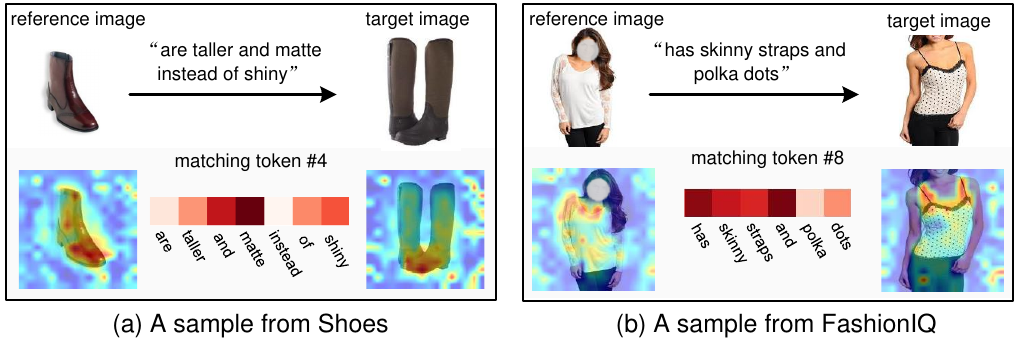}
    \vspace{-1.5em}
	\caption{Attention visualization for latent factor token learning.}
	\label{fig:attn_example}
	\vspace{-1.5em}
\end{figure}

\subsubsection{Generalization towards CIR}

To study the generality of the devised iterative dual \mbox{self-training} paradigm for the \mbox{CIR} task, 
we also selected the public \mbox{CLVC-Net}~\cite{clvcnet} model as the CIR model in our \mbox{self-training} paradigm.  We named the self-training boosted  \mbox{CLVC-Net} as  \mbox{CLVC-Net+}.
The experimental settings are identical to LIMN+.
Table~\ref{tab:baseline+} shows the performance of  \mbox{CLVC-Net} and  \mbox{CLVC-Net+} on the FashionIQ and Shoes dataset. As can be seen,  \mbox{CLVC-Net+} consistently outperforms the original \mbox{CLVC-Net}.
Overall, the results reflect the generalization ability of our proposed iterative dual self-training paradigm for the task of CIR.

\subsection{Case Study (RQ4)}
\textcolor{black}{
\subsubsection{Attention Visualization for Latent Factor Learning}
We visualized the attention mechanism for latent factor token learning with two test samples in Figure~\ref{fig:attn_example}. For each sample, we visualized the attention distribution of a representative latent factor token over the multimodal query and that over the target image, based on the self-attention weights in the final layer of the masked Transformer. As shown in Figure~\ref{fig:attn_example}(a), for the latent factor token $\#4$, our model focuses on the vamp area of the shoes in the reference and target image, and the word ``matte'' in the modification text. This suggests that the latent factor token $\#4$ is likely to refer to the vamp material attribute. Regarding Figure~\ref{fig:attn_example}(b), for the latent factor token $\#8$, our model highlights the shoulders area of the shirt in both the reference image and the target image, as well as the phrase ``skinny straps'' in the modification text. This observation implies that the latent factor token $\#8$ is primarily related to the shoulder design of the garment. Overall, these observations suggest our proposed LIMN can effectively model the latent query-target matching factors.
}
\subsubsection{Composed Image Retrieval Visualization}
Figure~\ref{fig:case_study} illustrates several \mbox{CIR} results obtained by our LIMN and LIMN+ on three datasets. We reported the top $5$ retrieved images for the limited space. We utilized blue and green boxes to indicate the reference and target images, respectively. 
As shown in Figure~\ref{fig:case_study}(a), we observed that LIMN fails to rank the target image in the first place, while LIMN+ successfully raises the target image from the second place to the first place. The reason is that the desired item feature ``leopard'' is rare ($54$ related samples) in the original training set. 
Benefiting from more training samples ($132$ pseudo triplets with the modification demand of the ``leopard'' feature), the \mbox{self-training} boosted LIMN+ model can perform better on reasoning the rare features, thereby gaining better image retrieval performance. This is also reflected in the case of Figure~\ref{fig:case_study}(b) sampled from the CIRR dataset, where LIMN ranks the target image at the third place, while LIMN+ tops the target image.
Regarding cases in Figure~\ref{fig:case_study}(c) and Figure~\ref{fig:case_study}(d), the target images are ranked at the first place by both LIMN and LIMN+, yet we noticed that the top $5$ images retrieved by LIMN+ meet the multimodal query better than those retrieved by LIMN. This suggests the superiority of our designed iterative dual \mbox{self-training} paradigm in CIR again.
Meanwhile, we also noticed some failing examples (see Figure~\ref{fig:case_study}(e)) of our models, where the target
image is not retrieved within the top $5$ places by both models.
Looking into this case, we observed that all the top $5$ retrieved shirts of our models meet the requirement of the multimodal query. We attributed this phenomenon to the flaws of the dataset itself, where the \mbox{false-negative} images are not labeled as target images~\cite{artemis}. Overall, these observations show our models' effectiveness and practical value, and validate the benefit of introducing the iterative dual \mbox{self-training} paradigm.

\section{Conclusion and Future Work}
In this work, we first present a novel CLIP-Transformer based muLtI-factor Matching Network (LIMN) to explore the latent matching factors affecting the matching evaluation between the multimodal query and the target image. Moreover, we design an iterative dual \mbox{self-training} paradigm to take full use of the potential unlabeled \mbox{reference-target} image pairs in the dataset, so as to relieve the problem of limited data size. Extensive experiments are conducted on four public datasets, and the results prove the effectiveness of our proposed method. 
As expected, we observed that by considering the multiple \mbox{query-target} matching factors, the designed LIMN method demonstrates superior performance compared to existing efforts. Additionally, the proposed iterative dual \mbox{self-training} paradigm can further boost the image retrieval performance by excavating the value of the potential unlabeled \mbox{reference-target} image pairs. Overall, this work provides a novel approach along with complete experiments, which advances the research in the \mbox{CIR community}.

\textcolor{black}{Currently, our proposed model achieves better performance at the cost of higher model complexity. In the future, we will develop a more efficient CIR model with superior retrieval performance. In addition,}
in this work, we primarily employed the IDC method as an assistant to improve the performance of the CIR method. Moving forward, we intend to expand our method to solve the CIR task and the IDC task simultaneously within a unified framework, as the two tasks are highly correlated. Moreover, we plan to adapt our method to accommodate the multi-turn interactive image retrieval task, which is essential in multimodal dialogue systems.


\bibliographystyle{IEEEtran}

\bibliography{reference}
\vspace{-1em}
\begin{IEEEbiography}[{\includegraphics[width=1in,height=1.25in,clip,keepaspectratio]{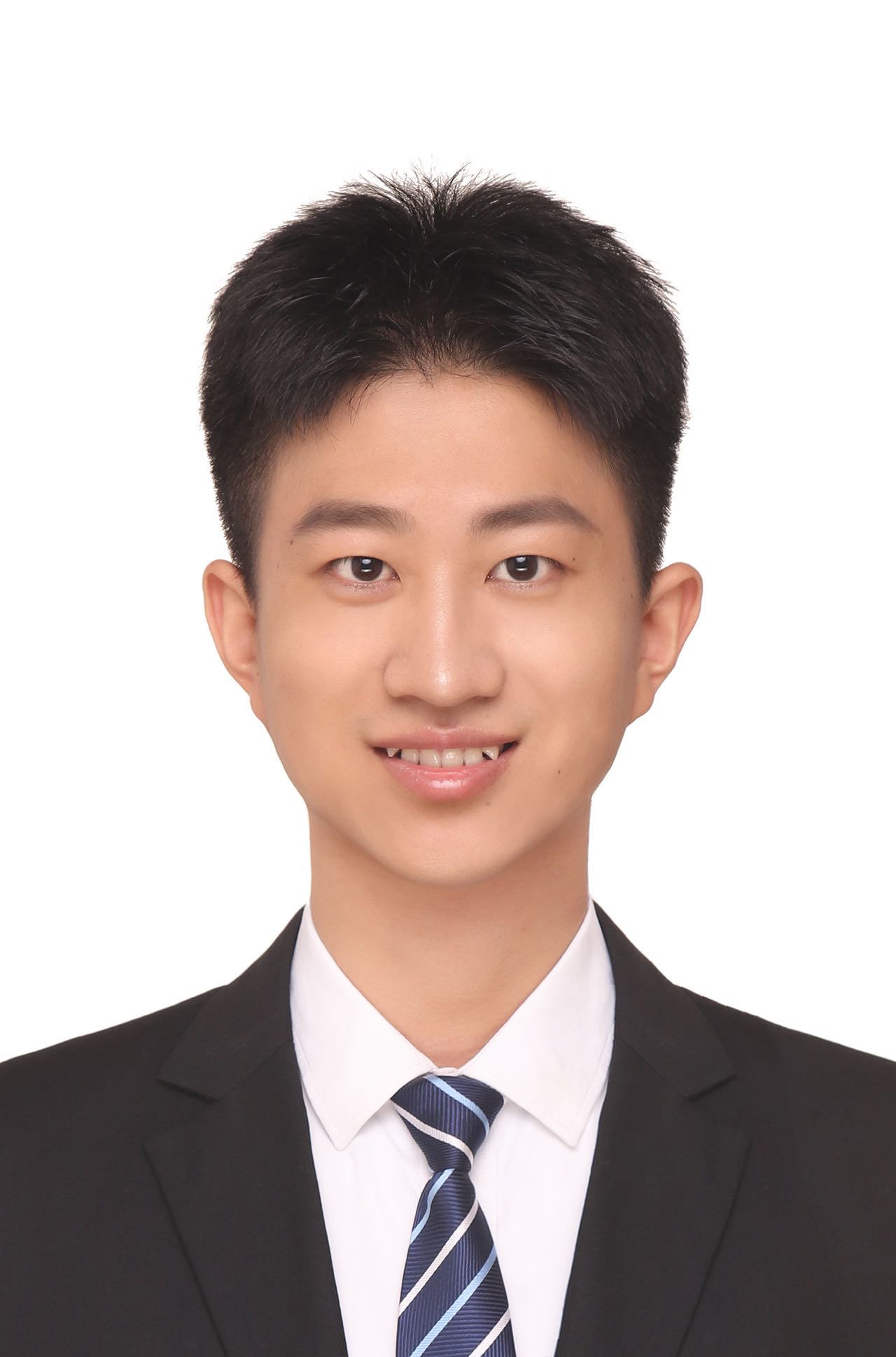}}]
{Haokun Wen} received the B.E. degree from the Ocean University of China, in 2019, and the master's degree from the School of Computer Science and Technology, Shandong University, in 2022. He is currently pursuing the Ph.D. degree with the Department of Computer Science and Technology, Harbin Institute of Technology (Shenzhen). His research interests include multimedia computing and information retrieval. He has published several papers in top venues, such as ACM SIGIR, ACM MM, and IEEE TIP.
\end{IEEEbiography}

\begin{IEEEbiography}[{\includegraphics[width=1in,height=1.25in,clip,keepaspectratio]{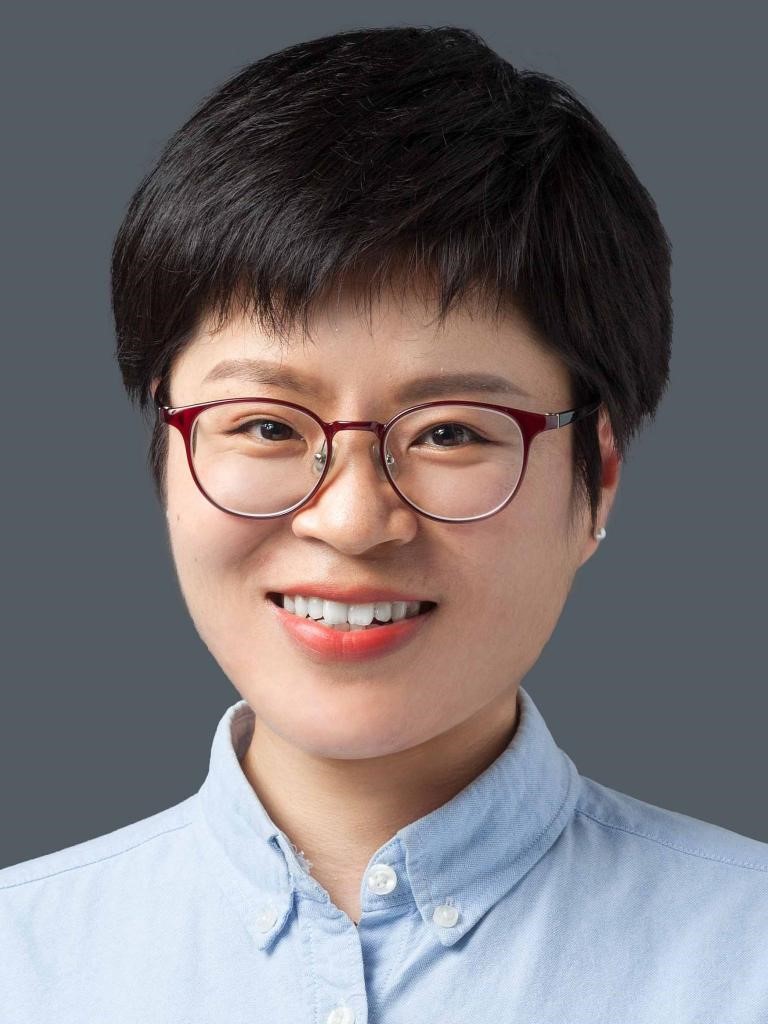}}]{Xuemeng Song}(Senior Member, IEEE)
	received the B.E. degree from the University of Science and Technology of China, in 2012, and the Ph.D. degree from the School of Computing, National University of Singapore, in 2016. She is currently an Associate Professor with Shandong University, China. She has published several papers in the top venues, such as ACM SIGIR, MM, and TOIS. Her research interests include information retrieval and social network analysis. She has served as a reviewer for many top conferences and journals. She is also an AE of IET Image Processing. 
\end{IEEEbiography}

\begin{IEEEbiography}[{\includegraphics[width=1in,height=1.25in,clip,keepaspectratio]{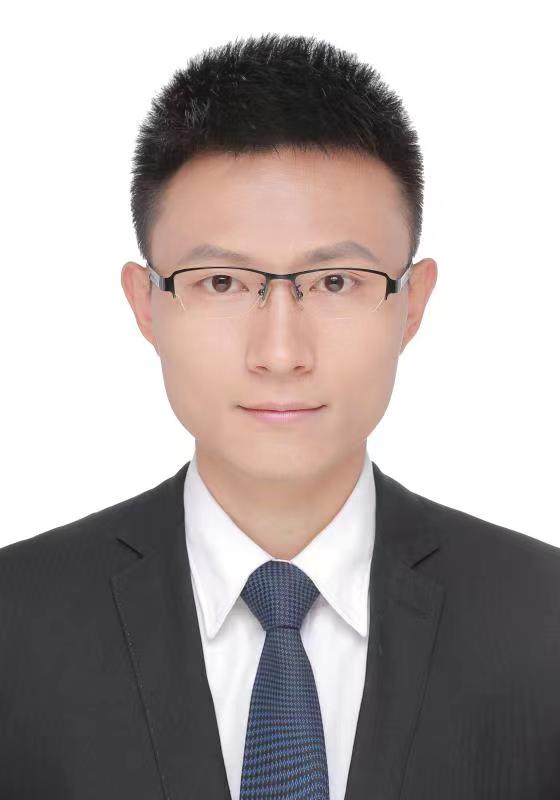}}]{Jianhua Yin}(Member, IEEE) received the Ph.D. degree in computer science and technology from Tsinghua University, Beijing, China, in 2017. He is currently an Associate Professor with the School of Computer Science and Technology, Shandong University, Jinan, China. He has published several papers in the top venues, such as ACM TOIS, IEEE TKDE, ACM MM, ACM SIGKDD, ACM SIGIR, and IEEE ICDE. His research interests mainly include data mining and machine learning applications.
\end{IEEEbiography}

\begin{IEEEbiography}[{\includegraphics[width=1in,height=1.25in,clip,keepaspectratio]{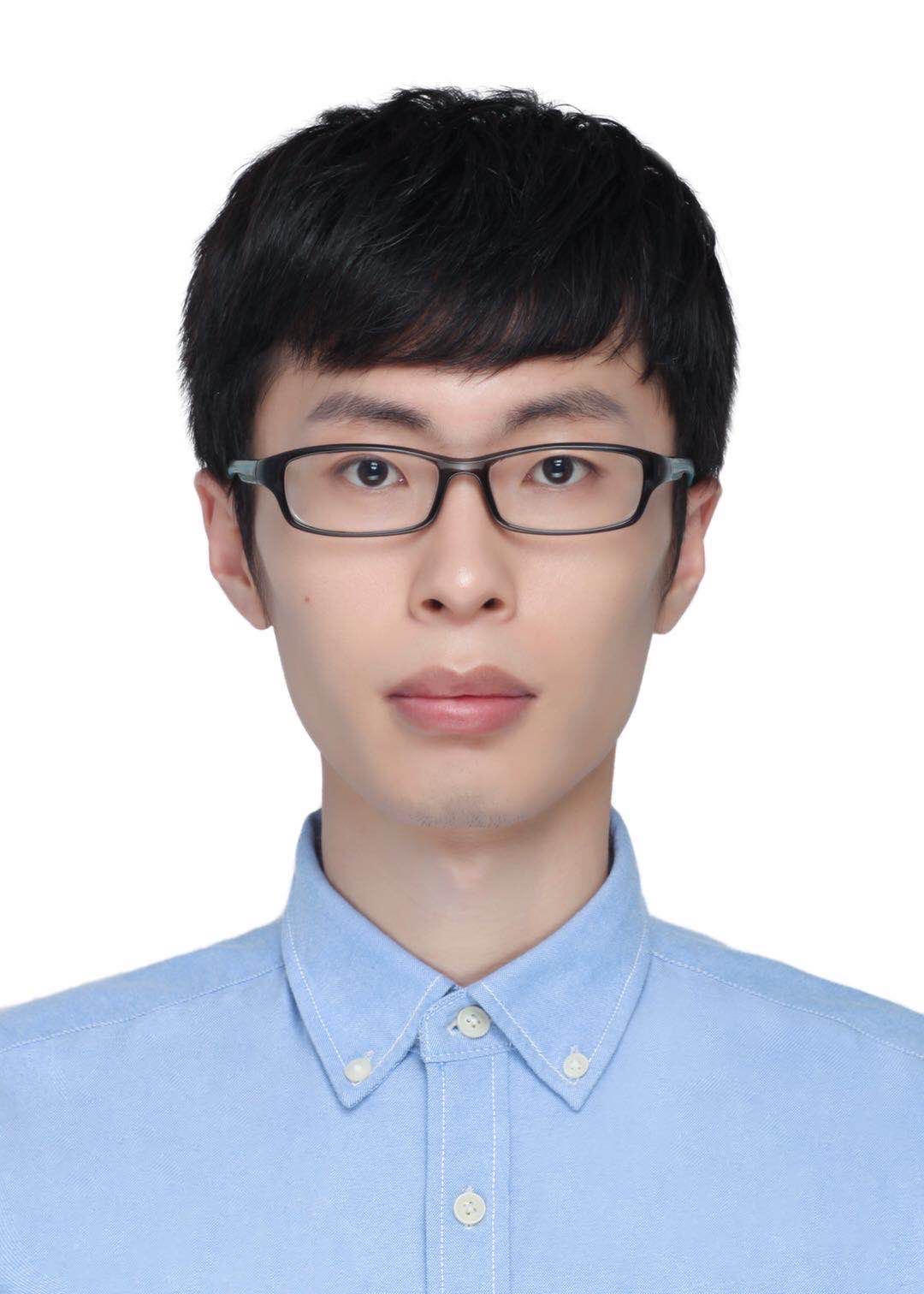}}]{Jianlong Wu}(Member, IEEE) received his B.Eng. and Ph.D. degree from Huazhong University of Science and Technology in 2014 and Peking University in 2019, respectively. He is currently an associate professor with Harbin Institute of Technology (Shenzhen). His research interests lie primarily in computer vision and machine learning. He has published more than 30 papers in top journals and conferences, such as TIP, ICML, NeurIPS, and ICCV. He serves as a Senior Program Committee Member of IJCAI 2021, an area chair of ICPR 2022/2020, and a reviewer for many top journals and conferences, including TPAMI, IJCV, ICML, and CVPR. He received many awards, such as the outstanding reviewer of ICML 2020, and the Best Student Paper of SIGIR 2021.
\end{IEEEbiography}

\begin{IEEEbiography}[{\includegraphics[width=1in,height=1.25in,clip,keepaspectratio]{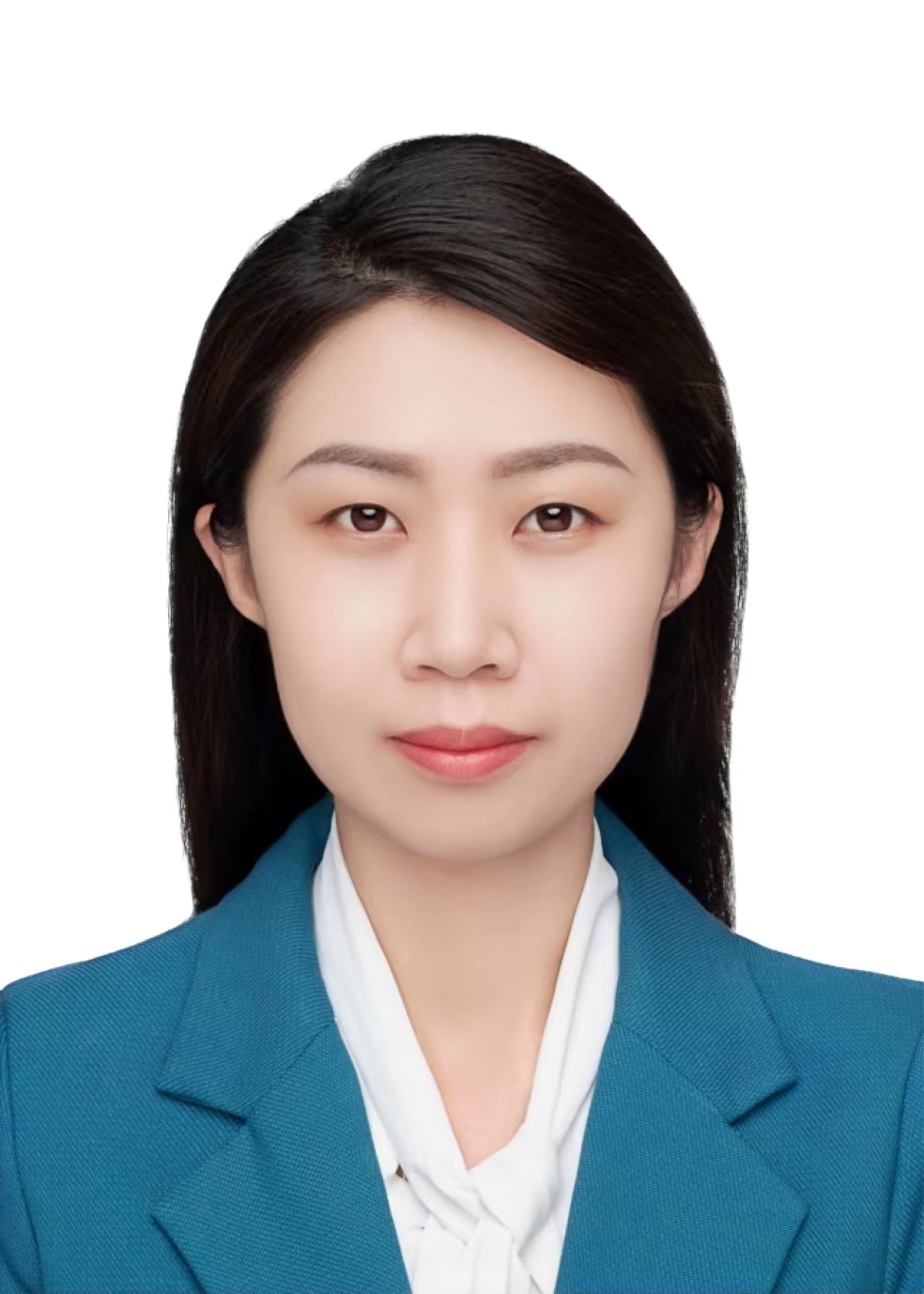}}]{Weili Guan}(Member, IEEE) received the master’s degree from National University of Singapore, and the Ph.D. degree from Monash University. She joined Hewlett Packard Enterprise in Singapore as a Software Engineer/Project Manager and worked there for around five years. She is currently a professor at the School of Electronics and Information Engineering, Harbin Institute of Technology (Shenzhen), China. Her research interests are multimedia computing and information retrieval. She has published more than 40 papers at the first-tier conferences and journals, like ACM MM, SIGIR, IEEE TPAMI and IEEE TIP.
\end{IEEEbiography}

\begin{IEEEbiography}[{\includegraphics[width=1in,height=1.25in,clip,keepaspectratio]{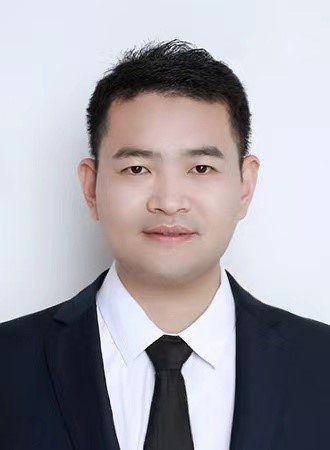}}]{Liqiang Nie}(Senior Member, IEEE) is currently the dean with the Department of Computer Science and Technology, Harbin Institute of Technology (Shenzhen). He received his B.Eng. and Ph.D. degree from Xi'an Jiaotong University and National University of Singapore (NUS), respectively. His research interests lie primarily in multimedia computing and information retrieval. Dr. Nie has co-/authored more than 100 papers and 4 books, received more than 23,000 Google Scholar citations. He is an AE of IEEE TKDE, IEEE TMM, IEEE TCSVT, ACM ToMM, and Information Science. Meanwhile, he is the regular area chair of ACM MM, NeurIPS, IJCAI and AAAI. He is a member of ICME steering committee. He has received many awards, like ACM MM and SIGIR best paper honorable mention in 2019, SIGMM rising star in 2020, TR35 China 2020, DAMO Academy Young Fellow in 2020, SIGIR best student paper in 2021, ACM MM best paper in 2022.
\end{IEEEbiography}

\end{document}